\documentclass[final,5p,times,twocolumn]{elsarticle}
\usepackage{graphicx}
\usepackage{url}
\usepackage{hyperref}

\usepackage{subfigure}
\usepackage{atlasphysics}

\journal{Physics Letters B}

\newcommand{\sigmattbar}{\ensuremath{\sigma_{\ttbar}}}
\newcommand{\lumitot}{\mbox{35\,pb$^{-1}$}}

\newcommand{\pmasym}[2]{^{+#1}_{-#2}}
\newcommand{\ql}{q_{\mathrm{lepton}}}
\newcommand{\MET}{E_{T}^{\mathrm{miss}}}

\begin{document}

%\biboptions{sort&compress} 

%%%%%%%%%%%%%%%%%%%%%%%%%%%%%%%%%%%%%%%%%%%%%%%%%%%%%%%%%%%%%%%%%%%%%%%%%%%%%%
% Preamble
%%%%%%%%%%%%%%%%%%%%%%%%%%%%%%%%%%%%%%%%%%%%%%%%%%%%%%%%%%%%%%%%%%%%%%%%%%%%%%
\begin{frontmatter}

\title{\vspace{-0.5cm} \flushright{\normalsize{CERN-PH-EP-2011-201}} \\ \center{Measurement of the top quark pair production cross-section \\ with ATLAS in the single~lepton channel}}

\author{The ATLAS Collaboration\fnref{fn1}}

\fntext[fn1]{See Appendix for the list of collaboration members}

\begin{abstract}

  A measurement of the production cross-section for top quark pairs
  ($\ttbar$) in $pp$ collisions at $\sqrt{s}=7~\TeV$ is presented using
  data recorded with the ATLAS detector at the Large Hadron
  Collider. Events are selected in the single lepton topology by
  requiring an electron or muon, 
  large missing transverse momentum and at least three jets. 
  With a data sample of 35 pb$^{-1}$, two different multivariate methods,
  one of which uses $b$-quark jet identification while the other does not,
  use kinematic variables to obtain cross-section measurements of
  $\sigmattbar = 187 \pm 11 \mathrm{(stat.)} ^{+18}_{-17}
  \mathrm{(syst.)} \pm 6 \mathrm{(lumi.)} \mathrm{\ pb}$
  and
  $\sigmattbar = 173 \pm 17 \mathrm{(stat.)} ^{+18}_{-16}
  \mathrm{(syst.)} \pm 6 \mathrm{(lumi.)} \mathrm{\ pb}$ respectively.
  The two measurements are in agreement with each other and with QCD calculations.
  The first measurement has a better a priori sensitivity and constitutes the main result of this Letter.
%  The former is combined with the ATLAS
%  measurement performed in the dilepton channel to give
%  $\sigmattbar = XX \pm X \mathrm{(stat.)} ^{+X}_{-X}
%  \mathrm{(syst.)} ^{+X}_{-X} \mathrm{(lumi.)} \mathrm{\ pb}$.

\end{abstract}

\begin{keyword}

  %% keywords here, in the form: keyword \sep keyword
  high-energy collider experiment \sep cross-section \sep top physics

  %% MSC codes here, in the form: \MSC code \sep code
  %% or \MSC[2008] code \sep code (2000 is the default)

\end{keyword}

\end{frontmatter}

%%%%%%%%%%%%%%%%%%%%%%%%%%%%%%%%%%%%%%%%%%%%%%%%%%%%%%%%%%%%%%%%%%%%%%%%%%%%%%%
% This is where the document really begins
%%%%%%%%%%%%%%%%%%%%%%%%%%%%%%%%%%%%%%%%%%%%%%%%%%%%%%%%%%%%%%%%%%%%%%%%%%%%%%%

\section{Introduction}
\label{s:intro}

Measurements of the production and decay properties of top quarks are of central importance to the 
Large Hadron Collider (LHC) physics programme.
Uncertainties on the theoretical predictions for the top quark pair production 
cross-section are now less than 10\%, and comparisons with experimental measurements 
%performed in different channels 
allow a precision test of the predictions of Quantum Chromodynamics. 
Furthermore, top quark pair production is an important background in many searches 
for physics beyond the Standard Model (SM).
New physics may also give rise
to additional $\ttbar$ production mechanisms or modifications of the
top quark decay channels, which can affect the measured $\ttbar$ cross-section.

In the SM the $\ttbar$ production cross-section in $pp$ collisions is calculated to be
165~$\pmasym{11}{16}$~pb~\cite{topxs1,topxs2,topxs3} 
at a centre-of-mass energy $\sqrt{s}=7~\TeV$, assuming a top quark mass of 172.5 GeV. 
Top quarks are predicted to decay to a $W$-boson and a $b$-quark ($t\rightarrow Wb$)
nearly 100\% of the time. 
Events with a $\ttbar$ pair can be classified as `single lepton', `dilepton', 
or `all hadronic' according to the decays of the two $W$-bosons: 
each can decay into quark-antiquark pairs ($W\rightarrow q_1 \bar{q_2}$) or a
lepton-neutrino pair ($W\rightarrow\ell\nu$). 
Events in the single lepton channel, 
when the lepton is an electron or a muon, 
are characterised by an isolated, prompt, energetic lepton, jets, and 
missing transverse momentum from the neutrino.
At the Tevatron the $\ttbar$ cross-sections at
$\sqrt{s}=1.8$ TeV and at $\sqrt{s}=1.96$ TeV have been measured by
CDF~\cite{cdfresult1,cdfresult2}
and D\O~\cite{d0result1,d0result2} in most channels.  
ATLAS and CMS have measured the $\ttbar$ cross-section at $\sqrt{s}=7$~TeV 
at the LHC~\cite{atlas3ipb,atlas35ipbdilepton,cmsdilepton,cmssemilepton}.

This Letter describes measurements of the $\ttbar$ cross-section in the
single lepton plus jets channel with $\lumitot$ of data recorded by ATLAS in 2010.
Taking advantage of the increased data sample, the measurement techniques
developed in Ref.~\cite{atlas3ipb} were extended to employ kinematic likelihood discriminants
to separate signal from background and measure the cross-section. 
Two multivariate methods, one that includes $b$-quark jet identification ($b$-tagging) 
and one which does not, 
use several variables each to discriminate $\ttbar$ events from the background. 
The two analyses are sensitive to different sources of systematic uncertainty. 
For instance, the analysis without $b$-tagging is more sensitive to the
multijet background, whereas the analysis with $b$-tagging is sensitive to the 
background from $W$-boson production in association with $b$- and $c$-quarks.
The clearer separation of signal and background
leads to a smaller statistical uncertainty for the analysis with $b$-tagging. 
Another significant difference between the two measurements is that
the analysis with $b$-tagging uses a profile likelihood that implements an {\it in situ} fit of
the dominant systematic uncertainties, which improves its performance considerably.

\section{The ATLAS detector}
\label{s:detdata}

    The ATLAS detector~\cite{atlasdet}  consists of an inner tracking
system (inner detector, or ID) surrounded by a thin 
superconducting solenoid providing a 2 T magnetic field, 
electromagnetic and hadronic calorimeters and a muon spectrometer (MS). 
The ID consists of silicon pixel and microstrip
detectors, surrounded by a transition radiation
tracker. The electromagnetic calorimeter is a 
lead/liquid-argon (LAr) detector. Hadron calorimetry is based
on two different detector technologies, with scintillator
tiles or LAr as active media, and with either steel, copper,
or tungsten as the absorber material. The MS includes
three large superconducting toroids arranged with an
eight-fold azimuthal coil symmetry around the calorimeters,
and a system of three stations of chambers for the
trigger and for track measurements. 
    
A three-level trigger system is used to select interesting events. The level-1 trigger is implemented in
hardware and uses a subset of detector information to reduce the event rate to a design value of at most
75 kHz. This is followed by two software-based trigger levels, level-2 and the event filter, which together
reduce the event rate to about 200 Hz which is recorded for analysis.

The nominal $pp$ interaction point at the centre of the detector 
is defined as the origin of a right-handed coordinate system. 
The positive $x$-axis is defined by the direction from the interaction
point to the centre of the LHC ring, with the positive 
$y$-axis pointing upwards, while the $z$-axis is along the beam direction. 
The azimuthal angle $\phi$ is measured around
the beam axis and the polar angle $\theta$ is the angle from the
$z$-axis. The pseudorapidity is defined as $\eta = -\ln \tan(\theta/2)$.

\section{Simulated event samples}\label{s:mc}
\label{mc.section}

Monte Carlo (MC) simulation was used for various aspects of the analysis. 
The simulation consists of an event generator interfaced to a parton shower 
and hadronisation model, the results of which are passed through
a full simulation of the ATLAS detector and trigger system \cite{atlsim,GEANT}.
MC simulation was used when data-driven techniques were not available 
or to evaluate relatively small backgrounds and certain sources of systematic uncertainty.

For the calculation of the acceptance of the $\ttbar$ signal the next-to-leading order 
(NLO) generator {\sc MC@NLO} v3.41 \cite{MC@NLO} 
was used with the top quark mass set to $172.5\GeV$ and with the 
NLO parton density function (PDF) set CTEQ66~\cite{cteq}.

$W$- and $Z$-boson production in association with jets was simulated with {\sc Alpgen}~v2.13, 
which implements the exact leading order (LO) matrix elements for final states with up to six partons
and uses the `MLM' matching procedure to remove the overlaps between samples with $n$ 
and $n+1$ final state partons~\cite{alpgen}. 
The LO PDF set CTEQ6L1~\cite{cteq} was used to generate 
$W$+jets and $Z$+jets events with up to five partons.
Diboson, $WW$, $WZ$ and $ZZ$ events were generated with 
{\sc Herwig} \cite{herwig1,herwig2}.
Like the diboson production, single-top is also a relatively small background and 
is simulated using {\sc MC@NLO}, invoking the `diagram
removal scheme'~\cite{diagrem} to remove overlaps between
single-top and $\ttbar$ final states.

Unless otherwise noted, all events were hadronised with {\sc Herwig},
using {\sc Jimmy} \cite{Jimmy} for the underlying event model. Details of the generator
and underlying event tunes used are given in Ref.~\cite{UEtunes}. 

\subsection{Systematic uncertainties on signal and background modelling}
\label{s:mcsyst}

The use of simulated $\ttbar$ samples to calculate the signal acceptance 
gives rise to various sources of systematic uncertainty.
These arise from the choice of the event generator and PDF set, and from
the modelling of initial and final state radiation (ISR and FSR).
The uncertainties due to the choice of generator and parton shower model 
were evaluated by comparing the results obtained with {\sc MC@NLO}
to those of {\sc Powheg}~\cite{powheg}, with events hadronised with either 
{\sc Herwig} or {\sc Pythia}~\cite{pythia}. 
The uncertainty due to the modelling of ISR/FSR was evaluated
using the {\sc AcerMC} generator~\cite{Acer} interfaced to
{\sc Pythia} and by varying the parameters controlling the ISR/FSR emission
by a factor of two up and down.
The variation ranges  used are comparable to those in 
\cite{Perugia} for ISR and ~\cite{Perugiabis} for FSR.  
Finally, the uncertainty in the PDF set used to generate $\ttbar$ samples 
was evaluated using a range of current
PDF sets with the procedure described in Ref.~\cite{PDF4LHC, Ball:2010de, Martin:2009iq}.

The production of the $W$+jets background based on MC simulation has
uncertainties on the total cross-section, on the contribution of
events with jets from heavy-flavour ($b,c$) quarks, and on the shape
of kinematic distributions. The predictions of the total cross-section
have uncertainties of order 50\%~\cite{wjetstheo}, increasing with
jet multiplicity. Total $W$+jets cross-section predictions were not used
in the cross-section measurement as this background was extracted from the
fit to the data (see Section~\ref{s:analysis}), but were used in the MC simulation
shown in Figs.~\ref{fig:3jetExclusive}~to~\ref{fig:5jetInclusive}. 
%The $W$+jets heavy-flavour content has been measured in $W$ events 
%with exactly two jets by studying the properties of secondary vertices inside 
%jets~\cite{WjetsHFcontent}. 
A combination of the fitting method described in \cite{WjetsHFcontent} and a counting method 
described here, both relying upon final states with one and two jets, was 
used to estimate the heavy flavour fractions in $W$+jets events.
Since these bins are dominated by
$W$+jet events, the total $W$+jet contribution to these events can be obtained, both with and without
requiring at least one $b$-tagged jet. These four numbers are then used to constrain the following four
event types which make up the $W$+jets sample: $W+b\bar{b}$, $W+c\bar{c}$, $W+c$ and $W$+light flavours. 
Additionally it was assumed that the k-factors for $W+b\bar{b}$ and $W+c\bar{c}$ are equal.
MC simulation with {\sc Alpgen} was used to estimate the $b$-tagging efficiencies for each sub-sample 
as well as to extrapolate from the one-jet to the two-jet bin. 
The dominant uncertainties in this method arise from jet energy scale and $b$-tagging
uncertainties. As a result of this study, it was found that the $W+b\bar{b}$ and 
$W+c\bar{c}$ sub-samples of events in the {\sc Alpgen} MC simulation
were to be rescaled by 1.30$\pm$0.65, whereas $W+c$ events were rescaled by 1.0$\pm$0.4.
An additional 25\% relative uncertainty per jet bin was assigned 
to these flavour fractions when applied to the signal region based upon studies 
with {\sc Alpgen} MC simulation.

%Based on these results, the simulation predictions for $W$-bosons produced in association 
%with $b$- or $c$-quark pairs have been scaled by factors of 1.30$\pm$0.65 or 1.0$\pm$0.4, 
%respectively. 
The uncertainty on the shape of $W$+jets kinematic distributions was assessed 
by changing the factorisation and renormalisation scales by a factor of two up and down;
and by varying the minimum $\pT$ of the final state quarks and gluons from 10 to 25 GeV,
with 15 GeV being the default.

For the smaller backgrounds arising from $Z$+jets, single-top and diboson production, only
the overall normalisation uncertainties were considered, 
taken to be 30\% for $Z$+jets production, 
10\% for single-top production, determined from comparisons of MCFM \cite{MCFM} and 
MC@NLO predictions, and 5\% for diboson production, determined from MCFM studies of scale 
and PDF uncertainties.

\section{Object selection}
\label{sec:objectselections}

Single lepton $\ttbar$ events are characterised by the presence of an electron or muon, jets, 
and missing transverse momentum, which is an indicator of undetected neutrinos, in the
final state.
The events used in this analysis were triggered by single-lepton triggers. 
The electron trigger required a level-1 electromagnetic cluster in the calorimeter
with transverse momentum $\ET > 10 \GeV$. 
A more refined cluster selection was applied in the level-2 trigger,
and a match between the electromagnetic cluster and an ID track was required in the event filter. 
The muon trigger required a track with transverse momentum $\pT > 10 \GeV$ in the muon trigger chambers
at level-1, matched to a muon of $\pT > 13 \GeV$ reconstructed in the
precision chambers and combined with an ID track at the event filter.
%The trigger efficiency for leptons which pass the offline cuts 
%is maximal, having reached the plateau region.
%Leptons which pass the offline selection are well within the plateau region of the trigger turn-on curves.

The same object definition used for the previous $\ttbar$ cross-section measurement~\cite{atlas3ipb} 
was used in this analysis, except for more stringent electron selection criteria and 
ID track quality requirements for muons.
Electron candidates were defined as electromagnetic clusters consistent with 
the energy deposition of an electron in the calorimeters and with an associated 
well-measured track.
They were required to satisfy $\pT > 20$~GeV and 
$\mathrm{|\eta_{cluster}| <}$~2.47, where $\mathrm{\eta_{cluster}}$ is the
pseudorapidity of the cluster associated with the candidate. Candidates in the
barrel to endcap calorimeter transition region 1.37~$\mathrm{< |\eta_{cluster}|<}$~1.52 were excluded. 
%Also, in order to suppress the background from photon conversions, the track was required to have an 
%associated hit in the innermost active pixel layer.
Muon candidate tracks were reconstructed from track segments in the different layers of the muon
chambers. These segments were combined starting from the outermost layer, with a procedure
that takes material effects into account, and matched with tracks found in the inner detector.
The final candidates were refitted using the complete track information from both detector
systems and required to satisfy $\pT > 20$ GeV and $|\eta| < 2.5$.  

To further reduce background from leptons produced in heavy-flavour or in-flight hadron 
decays the selected leptons were required to be `isolated'.
For electrons the transverse momentum, $\ET$, deposited in the calorimeter cells
inside an isolation cone of size $\Delta R = \sqrt{ (\Delta \eta)^2 + (\Delta\phi)^2 } = 0.2$ 
around the electron position was corrected to take into account the leakage 
of the electron energy into this cone. The remaining $\ET$ was required to be less than 4~GeV. 
Muons were required to have a distance $\Delta R$ greater than 0.4 from any jet with $\pT >20 \GeV$,
which suppresses muons from heavy-flavour decays inside jets.
Furthermore, the calorimeter transverse momentum in a cone of size $\Delta R = 0.3$ around the muon direction 
was required to be less than 4 GeV, and the sum of track transverse momenta, other than the muon track, 
in a cone of size $\Delta R = 0.3$ was required to be less than 4 GeV.

Pure samples of prompt muons and electrons were obtained from $Z$-boson
events in the data and were used to correct the lepton trigger, 
and the reconstruction and selection efficiencies
in MC simulation to match those in the data.
The corrections were found to be small.

Jets were reconstructed~\cite{topoclusters} with the anti-$k_t$ algorithm~\cite{antikt1,antikt2} 
with radius parameter 0.4 from clusters of adjacent calorimeter cells.
If the closest object to an electron candidate (before the above electron isolation requirement)
was a jet within a distance $\Delta R < 0.2$, 
the jet was removed. % to avoid double-counting of electrons as jets.
The jet energy scale (JES) and its uncertainty were derived by combining information from test-beam
data, LHC collision data and simulation. The JES uncertainty was found to vary from 2\% to 7\% as a
function of jet $\pT$ and $\eta$~\cite{JES}. 

Jets arising from the hadronisation of $b$-quarks were identified using an
algorithm (JetProb) ~\cite{jetprobCONF} which
relies upon the transverse impact parameter $d_0$ of each track in
the jet: this is the distance of closest approach in the transverse $x$-$y$ 
plane of a track to the primary vertex.
It is signed with respect to the jet direction: the sign is positive if the track 
crosses the jet axis in front of the primary vertex, negative otherwise.
The signed impact parameter significance, $d_0/\sigma_{d_0}$, of each selected track 
is compared to a resolution function for prompt tracks, to assess the probability 
that the track originates from the primary vertex. 
Here, $\sigma_{d_0}$ is the uncertainty on $d_0$.
The individual track probabilities 
are then combined into a global probability that the jet originates from the primary vertex.
The simulated data were smeared to reproduce the resolution found in collision data.

The $b$-tagging efficiencies and mistag rates
were calibrated with data for a wide range of $b$-tagging efficiency requirements.
The efficiency was measured in a sample of jets containing muons, making use of the transverse 
momentum of the muon relative to the jet axis.
The mistag rates were measured on an inclusive jet sample with two methods,
one using the invariant mass spectrum of tracks associated to reconstructed secondary 
vertices to separate light- and heavy-flavour jets, 
and the other based on the fraction of secondary vertices in data with negative decay-length significance.
The results of these measurements were applied 
in the form of $\pT$-dependent scale factors to correct the $b$-tagging 
performance in simulation to match the data.
For a $b$-tagging efficiency around 50\%, the 
scale factor was found to be approximately 0.9 in all bins of jet $\pT$,
and the relative $b$-tagging efficiency uncertainty was found to range from 5\% to 14\% 
depending on the jet $\pT$~\cite{jetprobCONF}.
The mistag rate and mistag scale factors are approximately 1\% and 1.1, respectively,
in the jet $\pT$ region of interest, $20 < \pT < 100$ GeV.
The analysis including $b$-tagging used the probabilities returned by the JetProb algorithm
as a discriminating variable, as explained in Section~\ref{s:analysis}.

The reconstruction of the missing transverse momentum $\MET$~\cite{metCONF} 
was based upon the vector sum of the transverse momenta of the reconstructed 
objects (electrons, muons, jets) as well as the transverse energy deposited 
in calorimeter cells not associated with these objects.
The electrons, muons and jets were used in the $\MET$ calculation consistently 
with the definitions and uncertainties stated above.

\section{Event selection}
\label{sec:eventselection}

Events that passed the trigger selection were required to contain exactly one reconstructed 
lepton with $\pT > 20 \GeV$, matching the corresponding event filter object.
Selected events were required to have at least one reconstructed primary
vertex with at least five tracks. Events were discarded if any jet with $\pT > 20 \GeV$
was identified to be due to calorimeter noise or
activity out of time with respect to the LHC beam crossings.
The $\MET$ was required to be greater than 35 (20) GeV in the electron (muon) channel and 
the transverse mass constructed from the lepton and $\MET$ transverse momentum vectors 
was required to be greater than 25 GeV (60 GeV $- \MET$) in the 
electron (muon) channel. The muon requirement is referred to as the `triangular cut'.
The requirements were stronger in the electron channel to suppress the larger multijet background.
Finally, events were required to have three or more jets with $\pT > 25 \GeV$ and $|\eta |<2.5$.
The selected events were then classified by the number of jets fulfilling 
these requirements and by the lepton flavour.
Table~\ref{tab:yields} shows the number of selected events in the data in the electron and muon channels, 
together with the SM expectations for the signal and the different backgrounds.
All predictions were obtained from MC simulation except the multijet background
estimate which was obtained from data as described in the next section.

\begin{table}[htb]
{\small
\caption{\small{
Number of observed events in the data in the electron and muon channels 
after the selection cuts as a function of the jet multiplicity. 
The expected signal and background contributions are also given. 
All simulated processes are normalized to theoretical SM predictions, 
except the multijet background which uses the normalisation
presented in Sec.~\ref{s:back}.
The quoted uncertainties include statistical, systematic and theoretical 
components, except for the multijet background.
All numbers correspond to an integrated luminosity of 35~$\ipb$.}
\label{tab:yields}}
\begin{center}
\begin{tabular}{l r@{ $\pm$ }l  r@{ $\pm$ }l  r@{ $\pm$ }l}
\hline
\hline
Electron channel &  \multicolumn{2}{c}{3 jets} &    \multicolumn{2}{c}{4 jets} &  \multicolumn{2}{c}{$\ge$ 5 jets}  \\
\hline
$\ttbar$         &   117 & 16                  &    109 & 15                   &   76  & 19   \\
$W$+jets         &   524 & 225                 &    124 & 77                   &   35  & 23   \\
Multijet         &   64  & 32                  &    12  & 6                    &   8   & 4    \\
Single top       &   21  & 5                   &    7   & 3                    &   3   & 2  \\
$Z$+jets         &   60  & 28                  &    21  & 15                   &   8   & 6    \\
Diboson          &    9  & 3                   &    1.9 & 1.5                  &   0.4 & 0.8  \\
\hline
Predicted        &   795 & 236                 &    275 & 84                   &   130 & 35   \\
Observed         &   \multicolumn{2}{c}{755}   &    \multicolumn{2}{c}{261}    &   \multicolumn{2}{c}{123}  \\ 
\hline
\hline
Muon channel     &  \multicolumn{2}{c}{3 jets} &    \multicolumn{2}{c}{4 jets} &   \multicolumn{2}{c}{$\ge$ 5 jets}   \\
\hline
$\ttbar$         &   165  & 22                 &    156 & 18                   &   108 & 27   \\
$W$+jets         &   976  & 414                &    222 & 139                  &   58  & 38   \\
Multijet         &   79   & 24                 &    18  & 6                    &   11  & 3    \\
Single top       &   31   & 7                  &    10  & 4                    &   4   & 2  \\
$Z$+jets         &   58   & 26                 &    14  & 10                   &   5   & 4    \\
Diboson          &   16   & 4                  &    3   & 2                    &   0.6 & 0.8  \\
\hline
Predicted        &   1325 & 422                &   423 & 143                   &   186 & 51   \\
Observed         &   \multicolumn{2}{c}{1289}  &   \multicolumn{2}{c}{436}     &   \multicolumn{2}{c}{190} \\ 
\hline
\hline
\end{tabular}
\end{center}
}
\end{table}

\newcommand{\nl} {N^{\mathrm{loose}}}
\newcommand{\nlr} {N^{\mathrm{loose}}_{\mathrm{prompt}}}
\newcommand{\nlf} {N^{\mathrm{loose}}_{\mathrm{non-prompt}}}
\newcommand{\nt} {N^{\mathrm{tight}}}
\newcommand{\ntr} {N^{\mathrm{tight}}_{\mathrm{prompt}}}
\newcommand{\ntf} {N^{\mathrm{tight}}_{\mathrm{non-prompt}}}
\newcommand{\epsr} {\ensuremath{\epsilon_{\mathrm{prompt}}}}
\newcommand{\epsf} {\ensuremath{\epsilon_{\mathrm{non-prompt}}}}

\section{Background evaluation}
\label{s:back}

The main backgrounds to $\ttbar$ signal events in the single lepton plus 
jets channel arise from $W$-boson production in association with jets, in
which the $W$ decays leptonically, and from multijet production. 
Smaller backgrounds arise from $Z$+jets, diboson and single-top 
production. These smaller backgrounds have been estimated 
from MC simulation and normalised to the latest theoretical predictions,
as discussed in Section~\ref{mc.section}.

The $W$+jets background is difficult to predict from theory, particularly in the high jet-multiplicity bins. 
A data-driven cross-check following methods similar to those described in Ref.~\cite{atlas3ipb} was therefore performed.
The results obtained with data were found to agree with the MC predictions within the uncertainties.
Both analyses presented here rely on the assumption that the MC simulation correctly 
describes the kinematic properties of the $W$+jets events,
whereas the normalisation of the $W$+jets cross-section was fitted from the data,
as described in Section \ref{s:analysis}.
In the analysis using $b$-tagging the theoretical uncertainty on the normalisation 
was used as a constraint in the fit, whereas in the
other analysis it was allowed to vary freely.

%The multijet background can arise from various sources:
%(i) jets mis-reconstructed as electrons if a relatively high fraction
%of their energy is in the electromagnetic calorimeter, 
%(ii) real electrons or muons produced in the decays of heavy-flavour hadrons inside jets, or
%(iii) photons inside jets which undergo conversions to electron-positron pairs.
%This background was measured with a data-driven approach.
The multijet background was measured with a data-driven approach.
In the muon channel, the background from multijet events is dominated by `non-prompt' muons arising 
from the decay of heavy-flavour hadrons, in contrast to the $\ttbar$ signal where
muons arise from the `prompt' decays of $W$-bosons.
The multijet background can be estimated by defining two samples of muons, `loose' and `tight'. 
The tight sample is the one defined in the event selection described above,
whilst the loose sample satisfy the same criteria {\it except} the muon isolation requirements.
Since the reconstructed muons from background are associated with jets, they
tend to be much less isolated than the leptons in $\ttbar$ decays. 
Any sample of muons is composed of prompt and non-prompt muons and it is assumed that the tight 
muon sample is a subsample of the loose sample:

\begin{eqnarray} 
\nl &=& \nlr + \nlf, \nonumber \\
\nt &=&  \epsr \nlr +  \epsf \nlf, 
\label{MM}
\end{eqnarray}

\noindent where $\nlf$ is the number of loose, non-prompt muons (with the other
$N^{x}_{y}$'s defined similarly) and $\epsr$ ($\epsf$) represents the probability for a 
prompt (non-prompt) muon that satisfies the loose criteria to also satisfy the tight ones. 
The probability $\epsr$ was measured from the data using high-purity samples dominated by $Z$-bosons 
decaying into muons. 
The probability $\epsf$ for a non-isolated lepton to pass the isolation cuts was measured
by defining control samples dominated by multijet events. 
Two different control samples were defined to have at least one jet plus a muon 
(i) with high impact parameter significance or
(ii) with low transverse mass of the muon-$\MET$ system plus reversed triangular cut.
These control samples gave consistent results. 
Contamination of the multijet control samples by muons from
$W$ and $Z$ events was determined from MC simulation. 
The results of these studies are $\epsf$ and $\epsr$ as a function of the
muon $\eta$, from which the multijet background expectations can be
obtained as a function of any variable.
A 30\% systematic uncertainty was assigned to this estimate based on the 
observation that the method gives agreement to within 30\% across
the different jet multiplicities.

In the electron channel, the multijet background also includes photons inside jets 
undergoing conversions into electron-positron pairs and jets with high electromagnetic fractions.
A different method was used, based on a binned likelihood fit of the $\MET$ distribution in the region $\MET < 35$ GeV.
The data was fitted to the sum of four templates:
multijet, $\ttbar$, $W$+jets and $Z$+jets.
The templates for the latter three processes were obtained from MC simulation whereas the multijet template was obtained
from the data in a control region defined by the full event selection criteria except that the
electron candidate fails one or more of the identification cuts.
The multijet background was obtained by extrapolating the fraction of multijet events from the fit at low 
$\MET$ to the signal region at high $\MET$.
Several choices of electron identification cuts were considered and the largest relative uncertainty among
these (50\%) was used as a conservative estimate of the systematic uncertainty of this background evaluation.

\section{Cross-section extraction}
\label{s:analysis}

The $\ttbar$ production cross-section was extracted by exploiting the kinematical
properties of $\ttbar$ events compared to those from the dominant background ($W$+jets)
by means of likelihood discriminants ($D$) constructed from several variables. 
Templates of the distributions $D$ for signal and all background samples 
were created using the TMVA package~\cite{tmva}.
The variables were selected for their good discriminating power, small correlation with
each other, and low sensitivity to potentially large uncertainties such as jet energy calibration.
The variables are:

\begin{itemize}

\item The pseudorapidity $\eta$ of the lepton, since leptons produced
in $\ttbar$ events are more central than those in $W$+jet events.

\item The aplanarity $\mathcal{A}$, defined as 3/2 times the smallest eigenvalue 
of the momentum tensor
$M_{ij} = { \sum_{k=1}^{N_\mathrm{objects}} p_{ik} p_{jk} }/
{ \sum_{k=1}^{N_\mathrm{objects}} p_k^2 }$, 
where $p_{ik}$ is the $i$-th momentum component of the $k$-th object
and $p_k$ is the modulus of its momentum. 
The lepton and the four leading jets are the objects included in the sum.
To increase the separation power of the aplanarity distribution, 
the transformed variable $\exp{(-8\times\, \mathcal{A})}$ was used. 
This variable exploits the fact that $\ttbar$ events are more isotropic than $W$+jets events. 

\item  The charge of the lepton $\ql$,
which uses the fact that a sample of $\ttbar$ events should contain
the same number of positively and negatively charged leptons,
while $W$+jet events produce an excess of positively charged leptons
in $pp$ collisions.

\item  $H_{\mathrm{T},3p}$, defined as the sum of the transverse energies of the third and fourth leading jets
normalised to the sum of the absolute values of the longitudinal momenta of the four leading jets, the lepton and
the neutrino,
$H_{\mathrm{T},3p} =  \sum_{i=3}^{4}|p_{\mathrm{T},i}^{\mathrm{jet}}|/
\sum_{j=1}^{N_\mathrm{objects}} |p_{z,j}|$,
where $\pT$ is the transverse momentum and $p_z$ the longitudinal momentum. 
The longitudinal momentum of the neutrino was obtained 
%by solving the event kinematics 
using the quadratic $W$ mass constraint and taking the solution with the
smaller neutrino $p_z$ value.
To increase the separation power of the $H_{\mathrm{T},3p}$ distribution, 
the transformed variable $\exp{(-4\times\,H_{\mathrm{T},3p})}$ was used.

\item The average ${\overline w}_{\mathrm{JP}}$ of $w_{\mathrm{JP}} = -\log_{10} P_l$ 
for the two jets with lowest $P_l$ in the event.
$P_l$ is the probability for a jet to be a light jet from the JetProb $b$-tagging algorithm.
These correspond to the jets that have the highest probability to be heavy-flavour jets.

\end{itemize}

Two complementary analyses were performed, one which relied upon the use 
of $b$-tagging information (i.e. the variable ${\overline w}_{\mathrm{JP}}$)
and one which did not. 
We refer to the analyses as `tagged' and `untagged', respectively.
The untagged analysis employed the first three variables, 
whereas the tagged analysis did not consider 
the lepton charge but used $H_{T,3p}$ and ${\overline w}_{\mathrm{JP}}$.
$\overline{w}_{\mathrm{JP}}$ was not included in the three-jet bin.
Figures \ref{fig:3jetExclusive}~to~\ref{fig:5jetInclusive} show the distributions 
of the discriminating variables for the selected data superimposed on the 
signal and background SM predictions for the different jet multiplicities.

\scriptsize{
\begin{figure}[tbp]
\begin{center}
\begin{tabular}{c}
\includegraphics[width=0.76\linewidth]{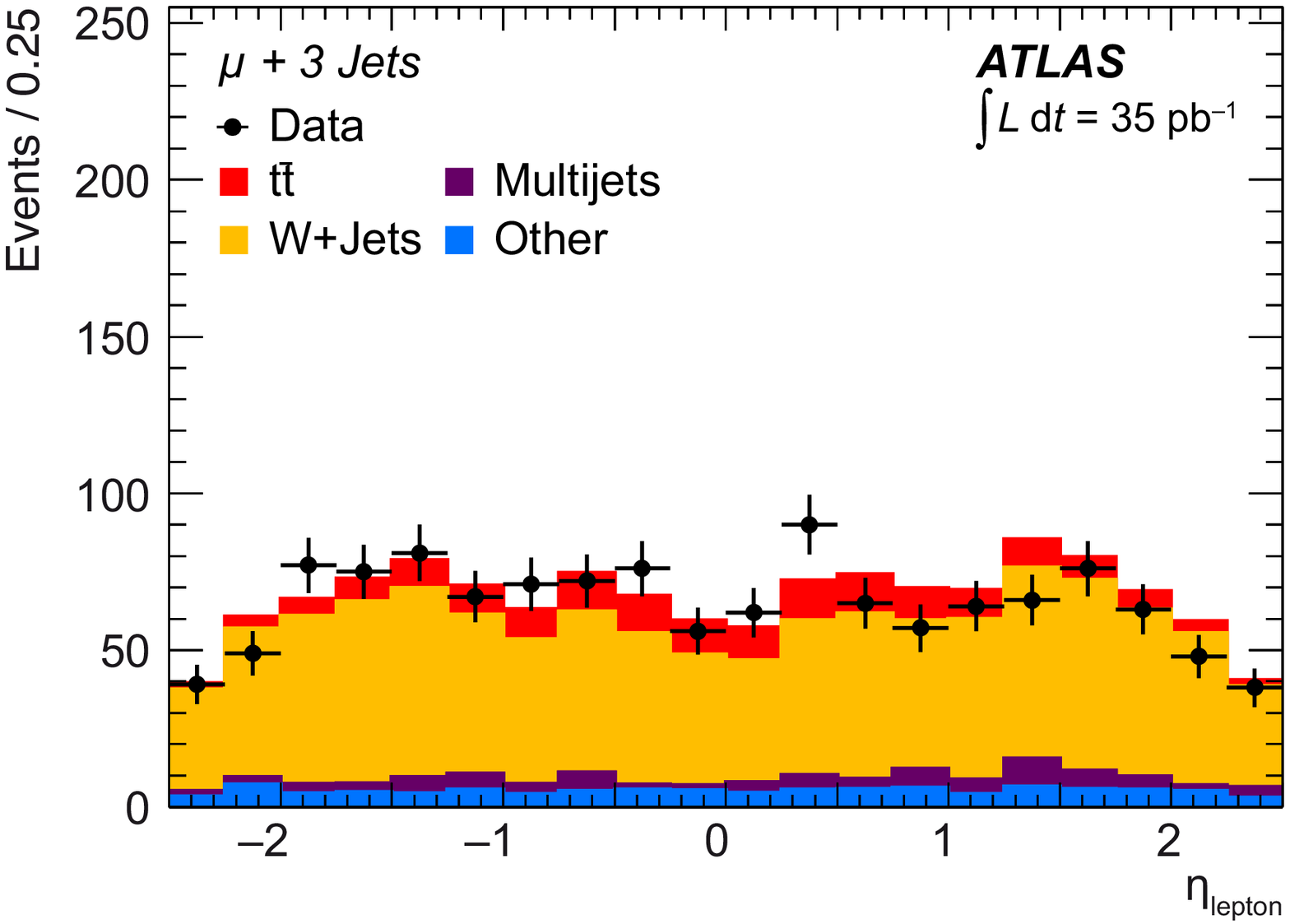} \\
\includegraphics[width=0.85\linewidth]{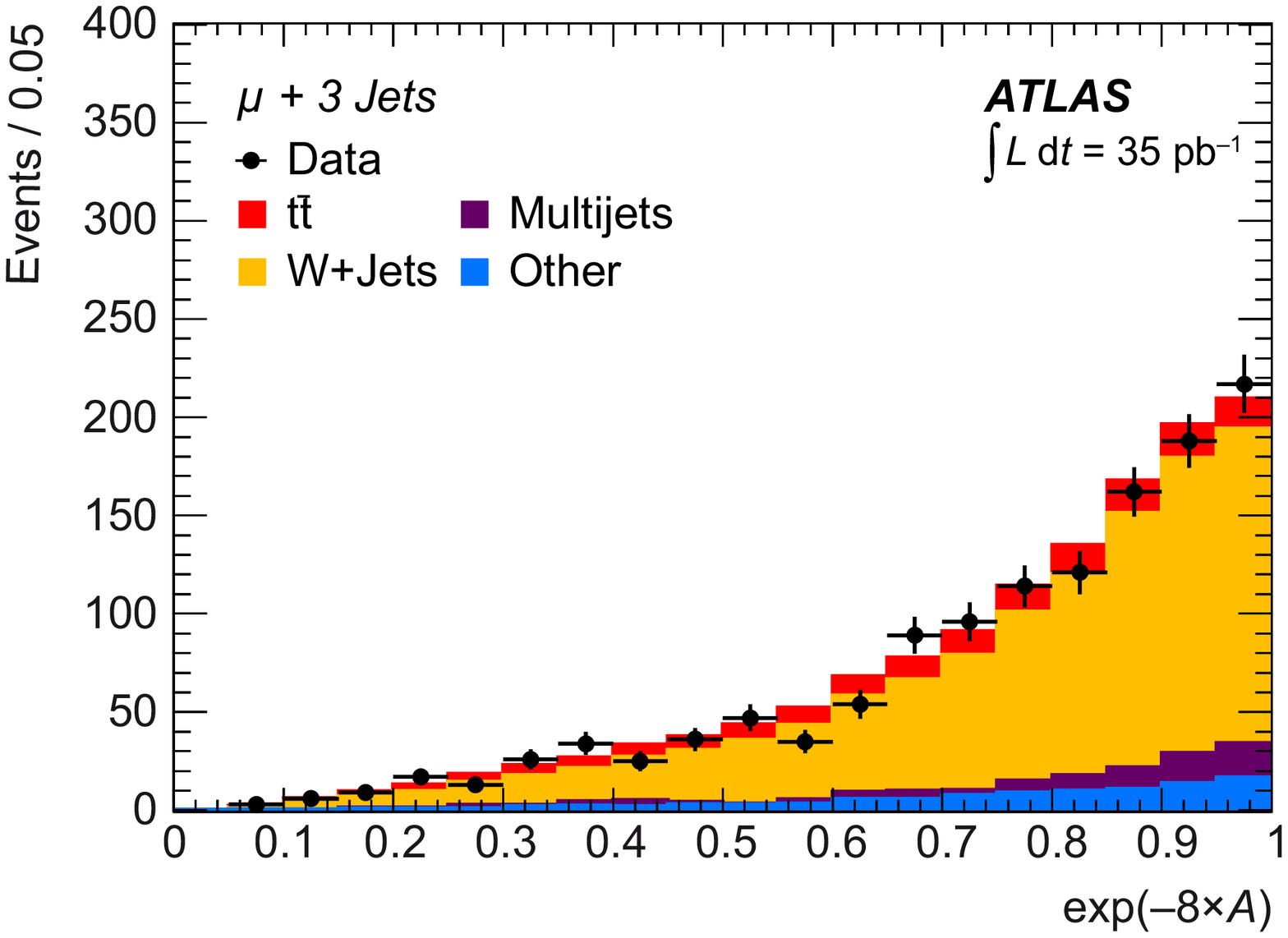} \\
\includegraphics[width=0.85\linewidth]{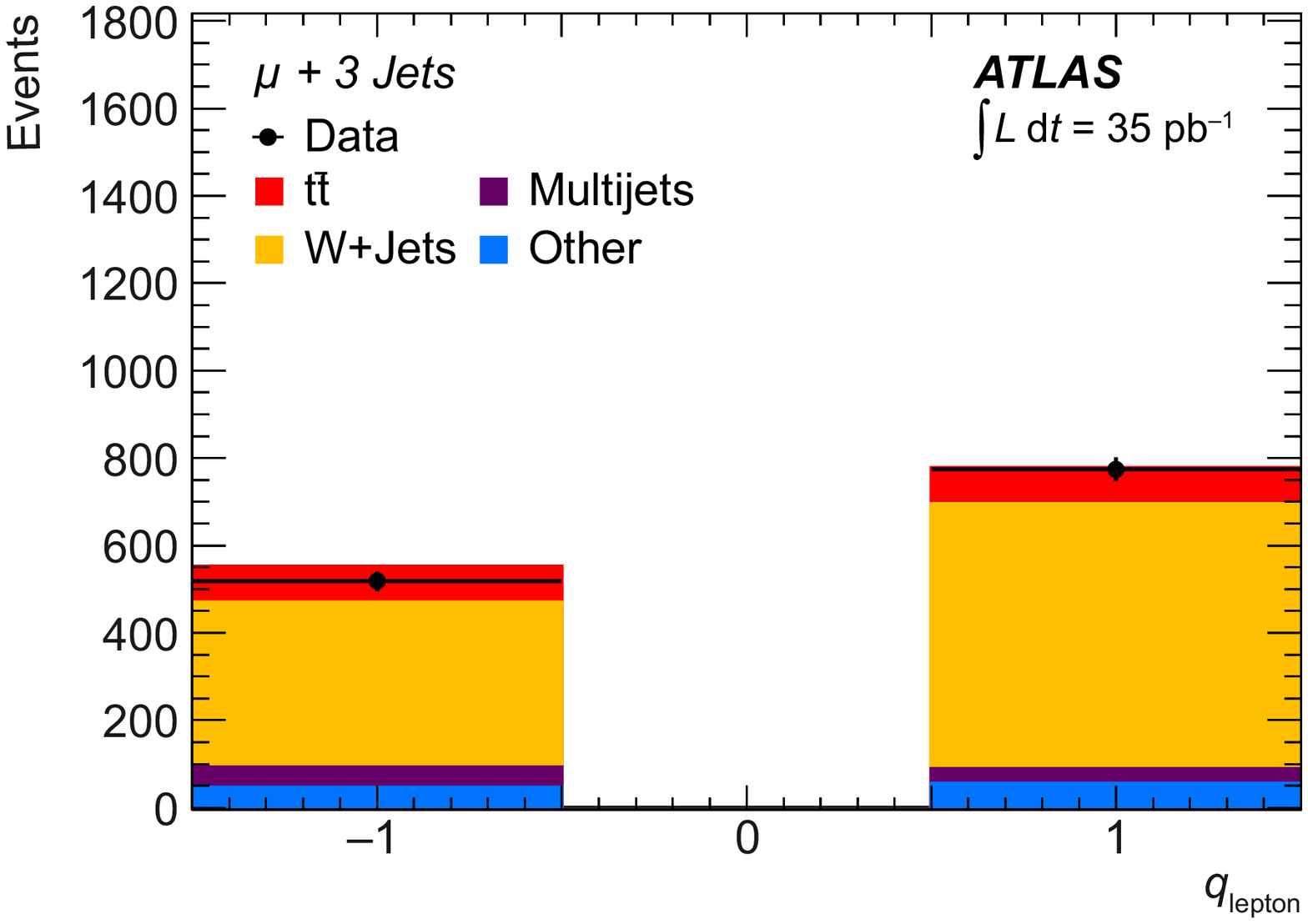} \\
\includegraphics[width=0.85\linewidth]{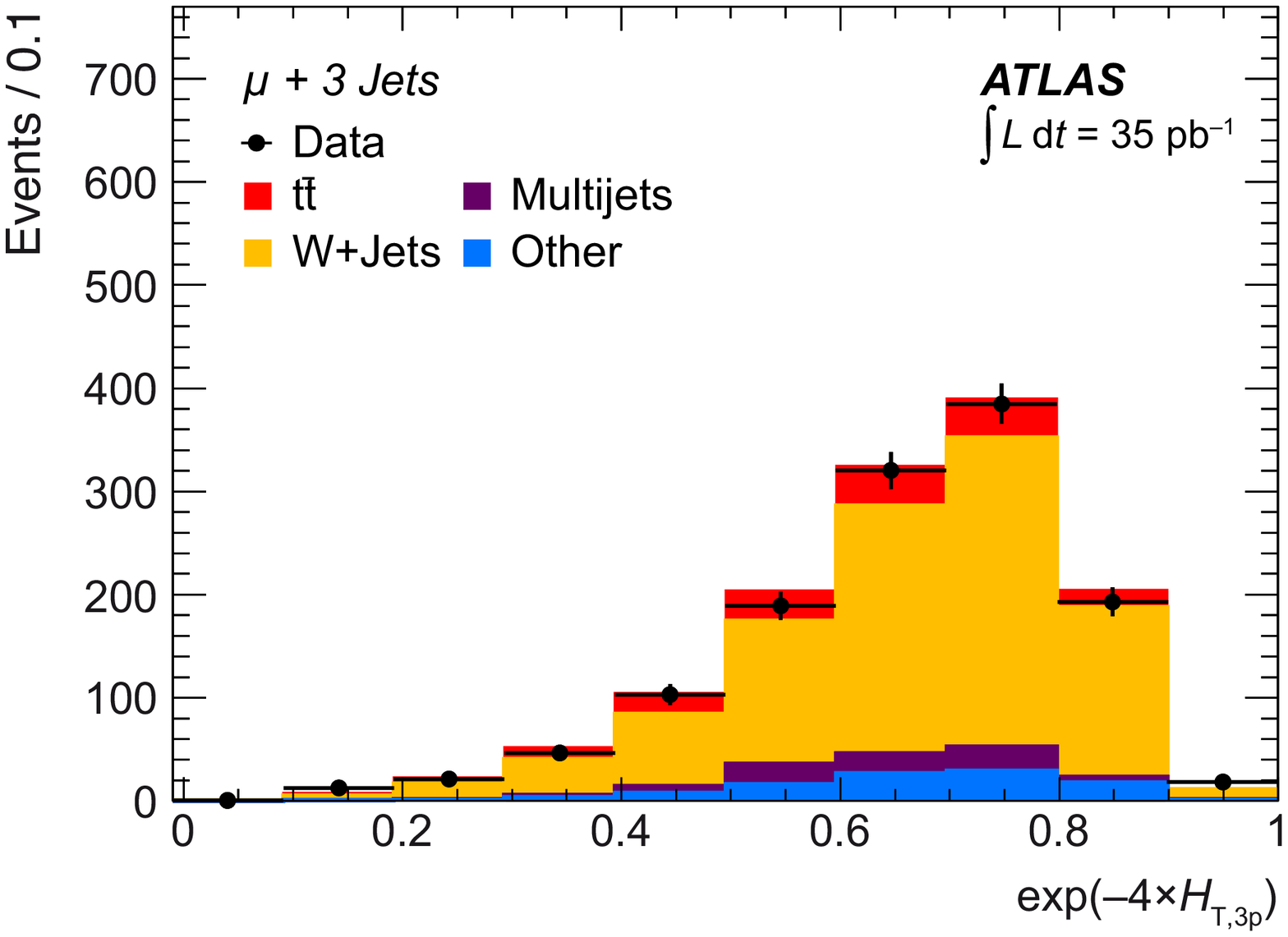} \\
\end{tabular}
\end{center}
\caption{\label{fig:3jetExclusive} 
Input variables to the likelihood discriminants in the exclusive three-jet bin for the muon channel:
lepton $\eta$ (top), 
$\exp(-8 \times \mathcal{A})$ (second from top),
lepton charge (third from top) and
$\exp(-4 \times H_{\mathrm{T},3p})$ (bottom).
All simulated processes are normalized to theoretical SM predictions, 
except the multijet background which uses the normalisation
presented in Sec.~\ref{s:back}.
The two top distributions are used in the untagged and the tagged analyses, 
the third distribution in the untagged analysis,
and the bottom distribution in the tagged analysis.
}
\end{figure}

\scriptsize{
\begin{figure}[tbp]
\begin{center}
\begin{tabular}{c}
\includegraphics[width=0.9\linewidth]{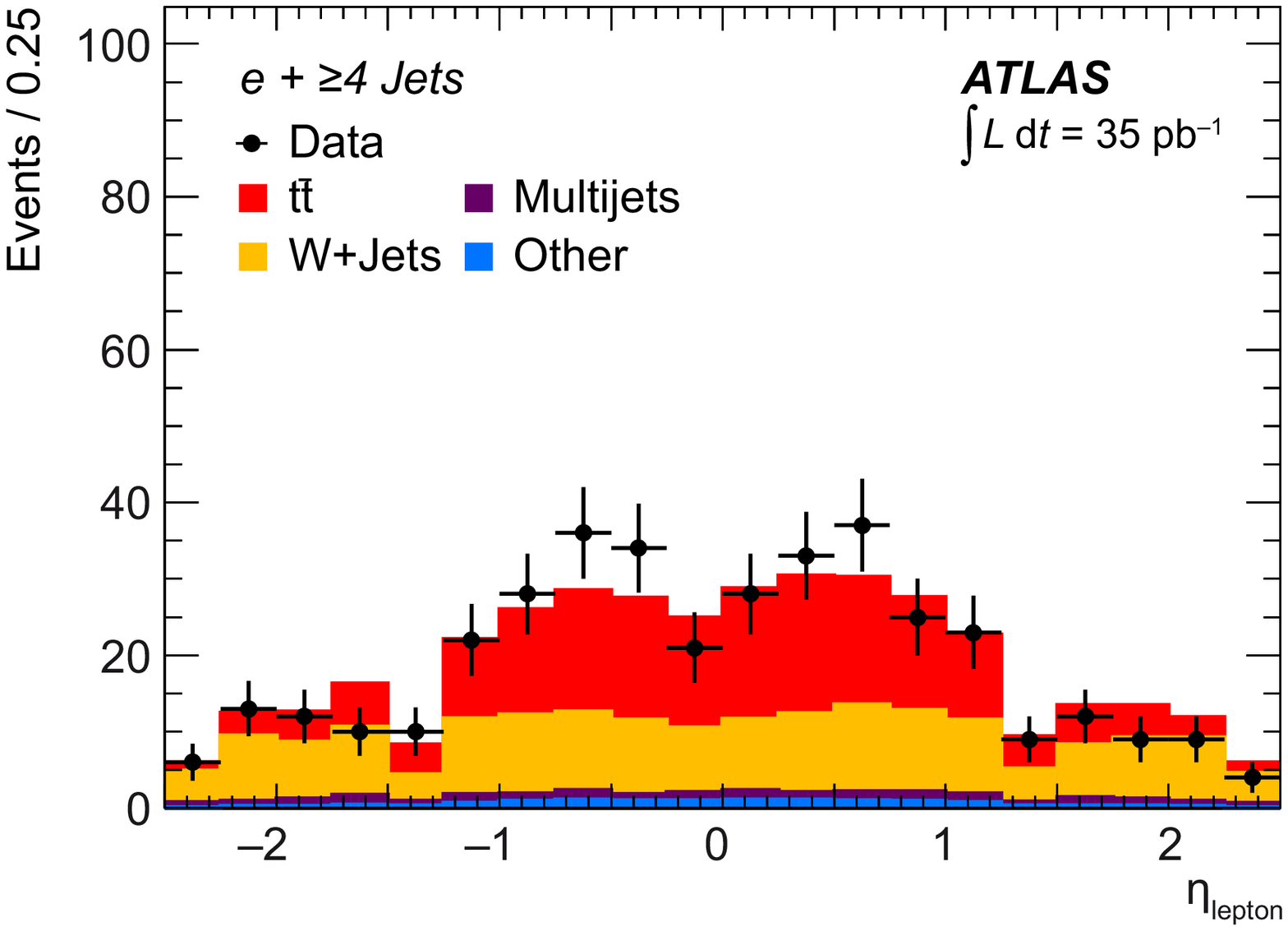} \\
\includegraphics[width=0.9\linewidth]{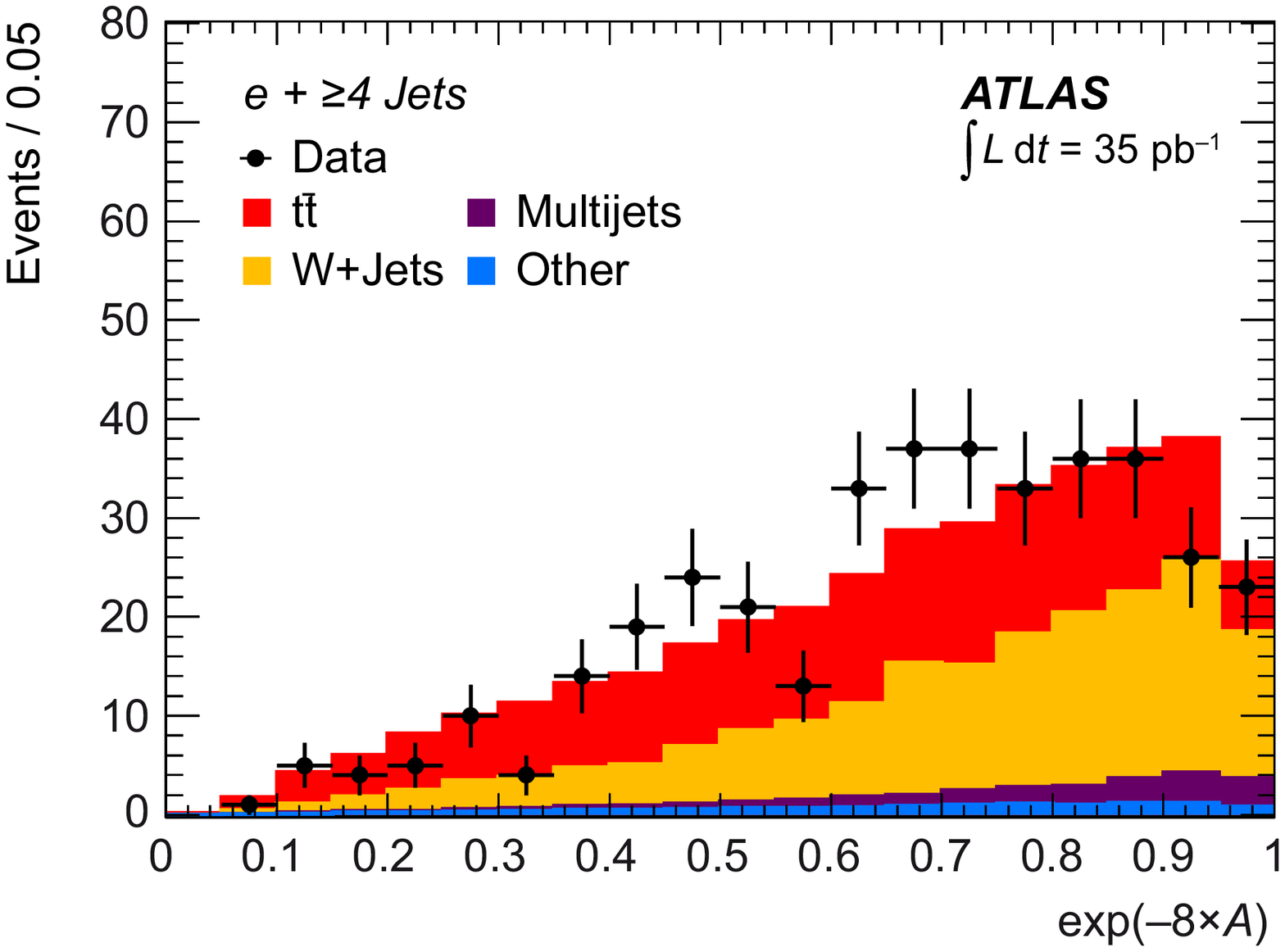} \\
\includegraphics[width=0.9\linewidth]{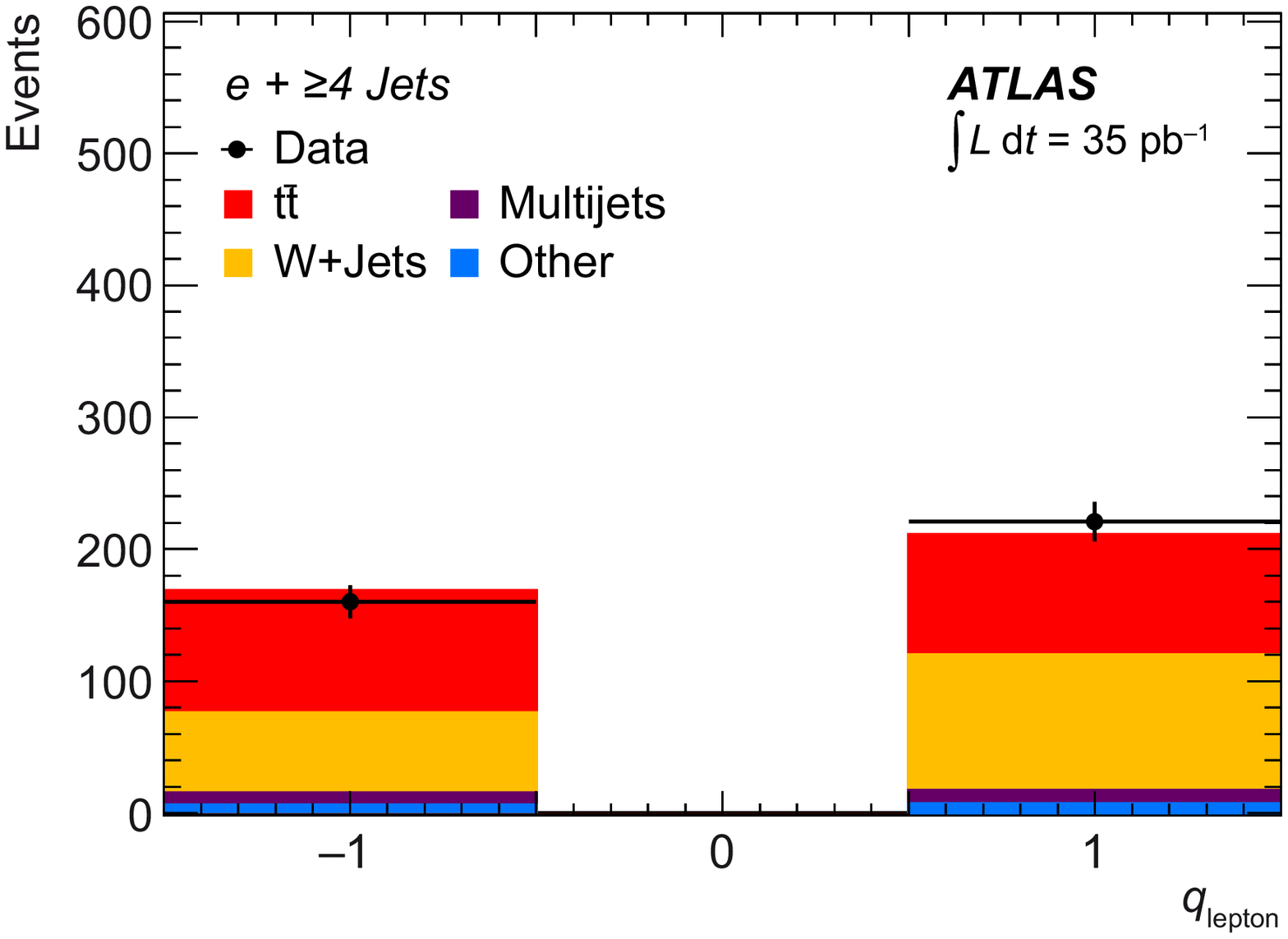} \\
\end{tabular}
\end{center}
\caption{\label{fig:4jetInclusive} 
Input variables to the likelihood discriminants in the inclusive four-jet bin for the electron channel:
lepton $\eta$ (top), 
$\exp(-8 \times \mathcal{A})$ (middle) and
lepton charge (bottom).
All simulated processes are normalized to theoretical SM predictions, 
except the multijet background which uses the normalisation
presented in Sec.~\ref{s:back}.
These distributions are used in the untagged analysis.
}
\end{figure}

\scriptsize{
\begin{figure}[tbp]
\begin{center}
\begin{tabular}{c}
\includegraphics[width=0.85\linewidth]{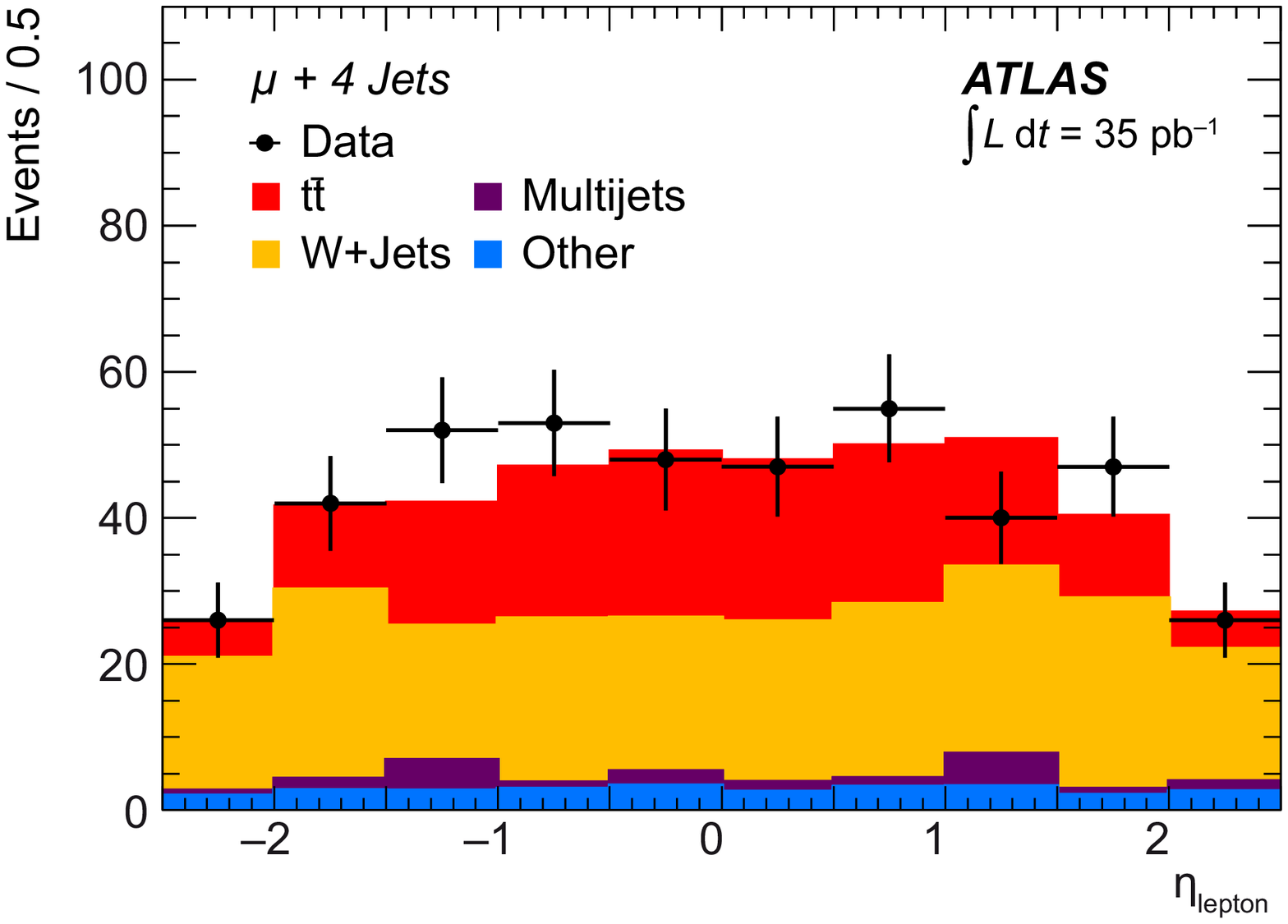} \\
\includegraphics[width=0.85\linewidth]{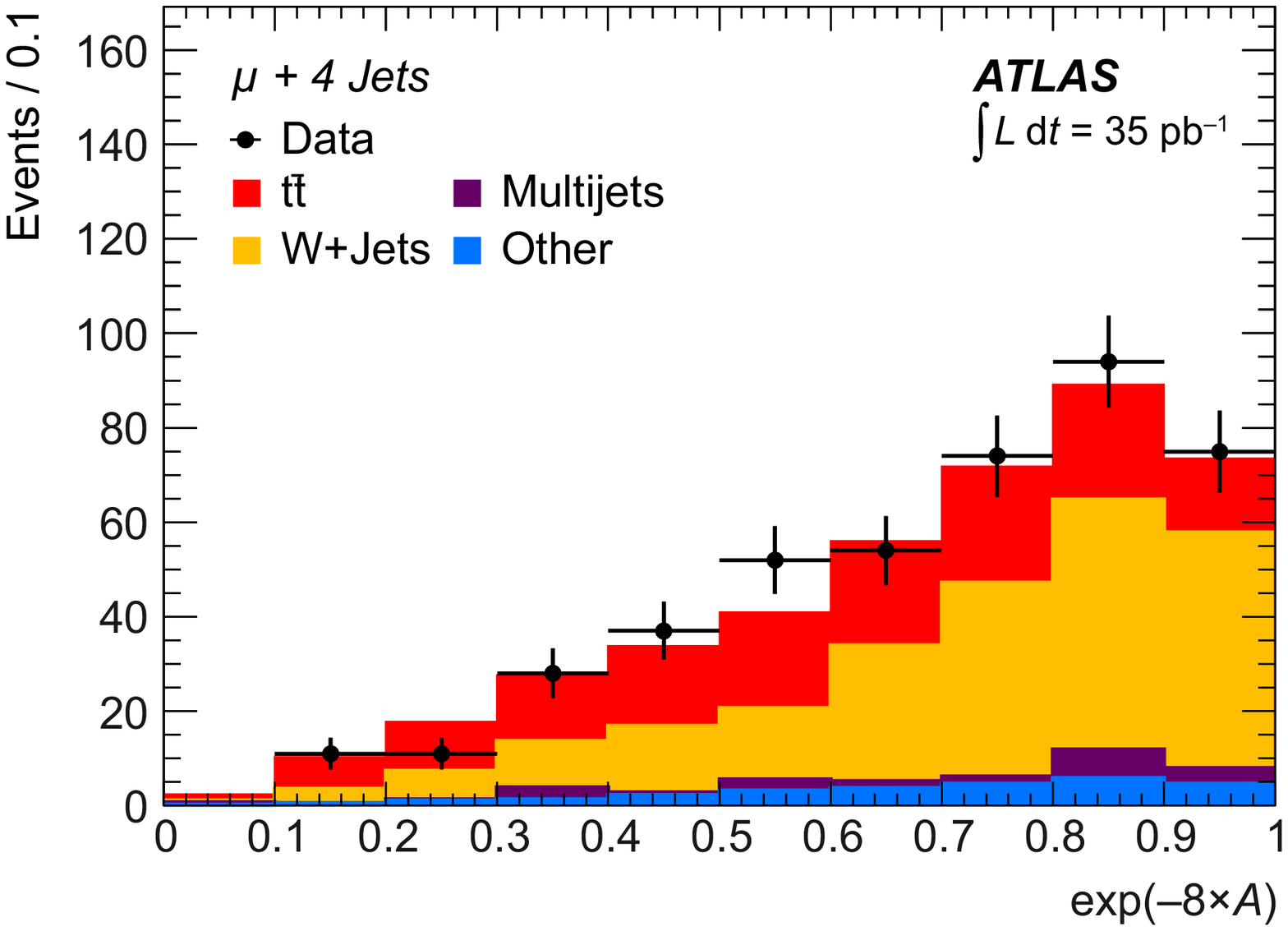} \\
\includegraphics[width=0.85\linewidth]{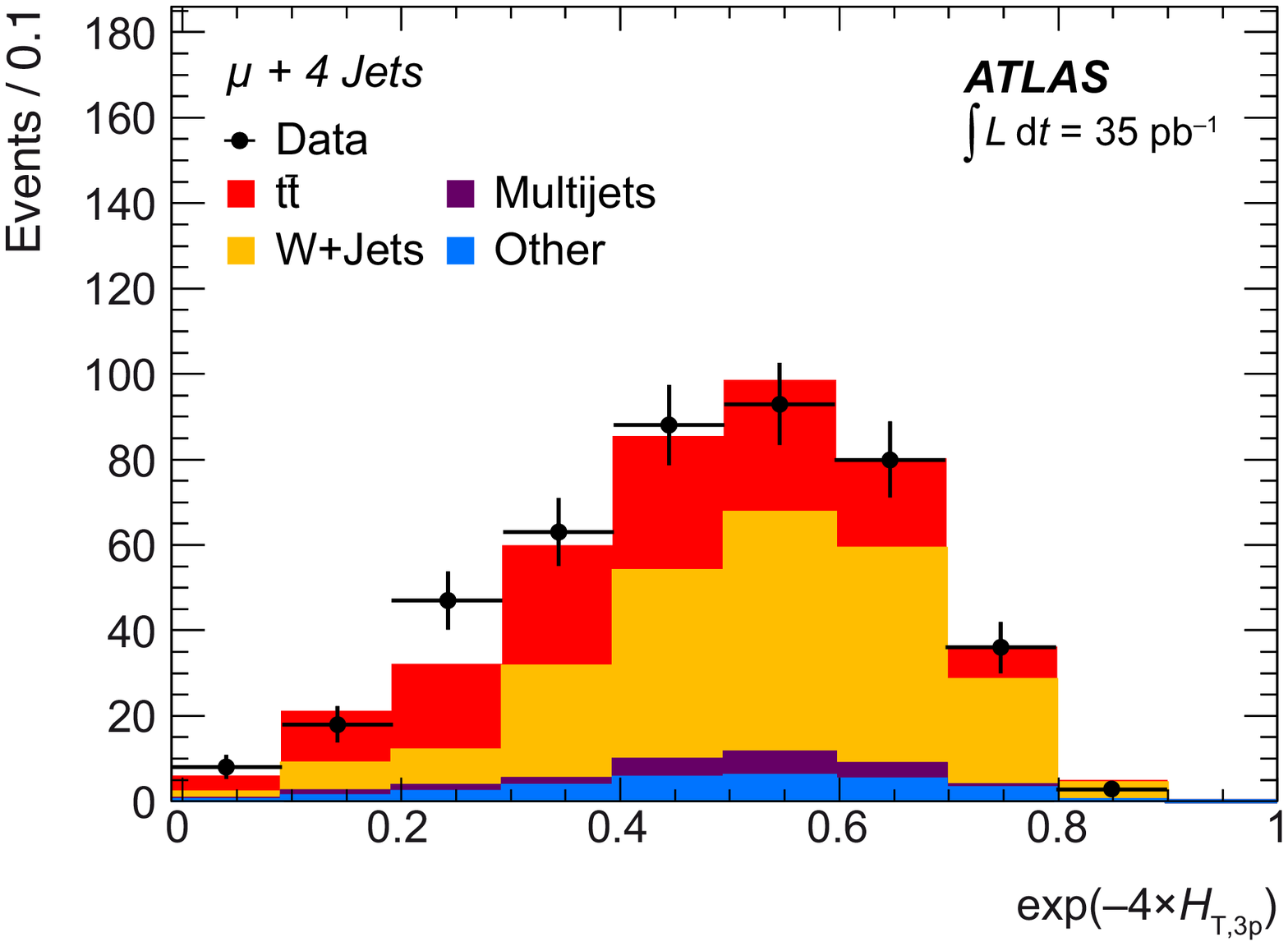} \\
\includegraphics[width=0.85\linewidth]{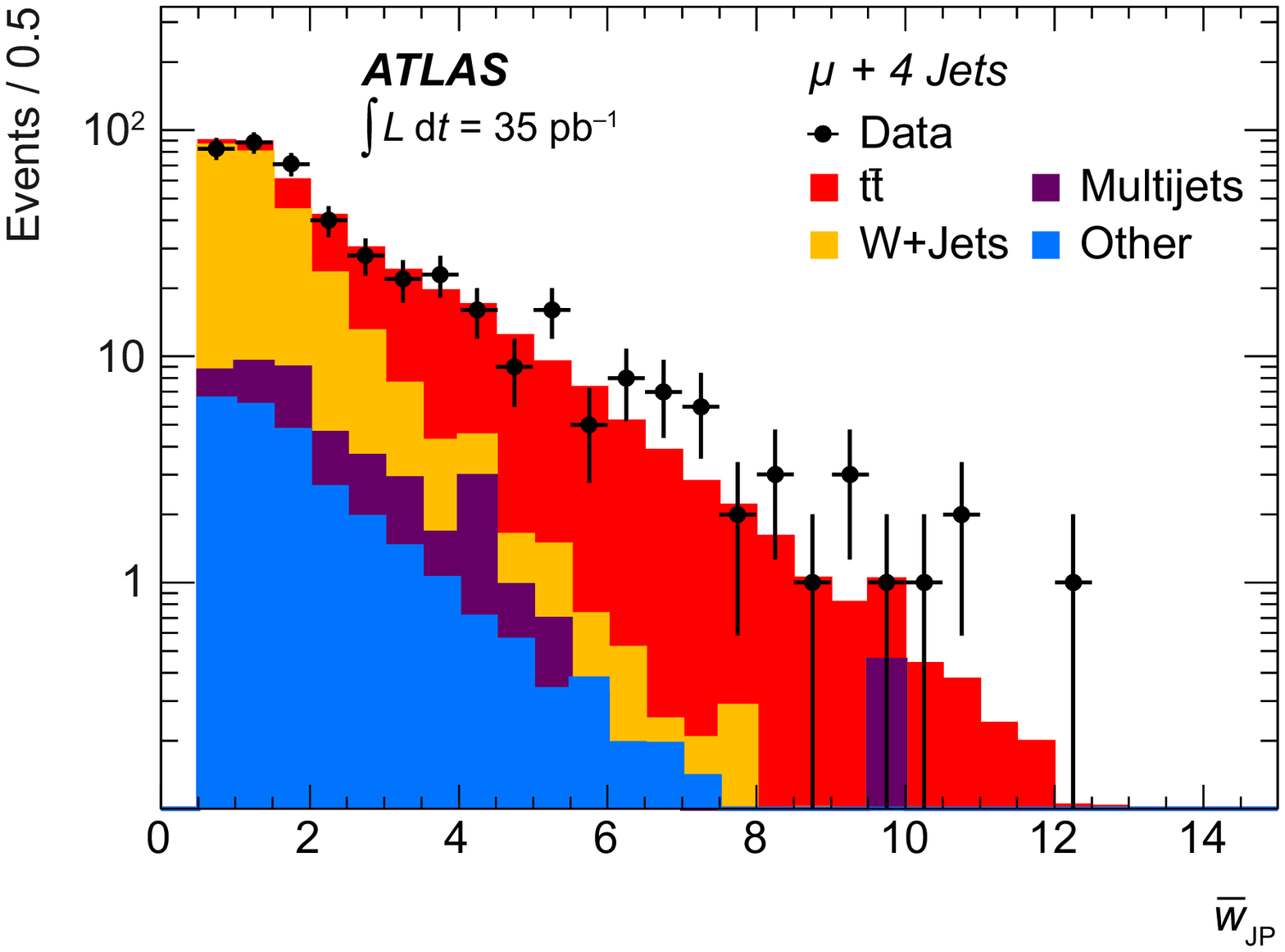} \\
\end{tabular}
\end{center}
\caption{\label{fig:4jetExclusive} 
Input variables to the likelihood discriminants in the exclusive four-jet bin for the muon channel:
lepton $\eta$ (top), 
$\exp(-8 \times \mathcal{A})$ (second from top),
$\exp(-4\times H_{\mathrm{T},3p})$ (third from top) and
${\overline w}_{\mathrm{JP}}$ (bottom).
All simulated processes are normalized to theoretical SM predictions, 
except the multijet background which uses the normalisation
presented in Sec.~\ref{s:back}.
These distributions are used in the tagged analysis.
}
\end{figure}

\scriptsize{
\begin{figure}[tbp]
\begin{center}
\begin{tabular}{c}
\includegraphics[width=0.85\linewidth]{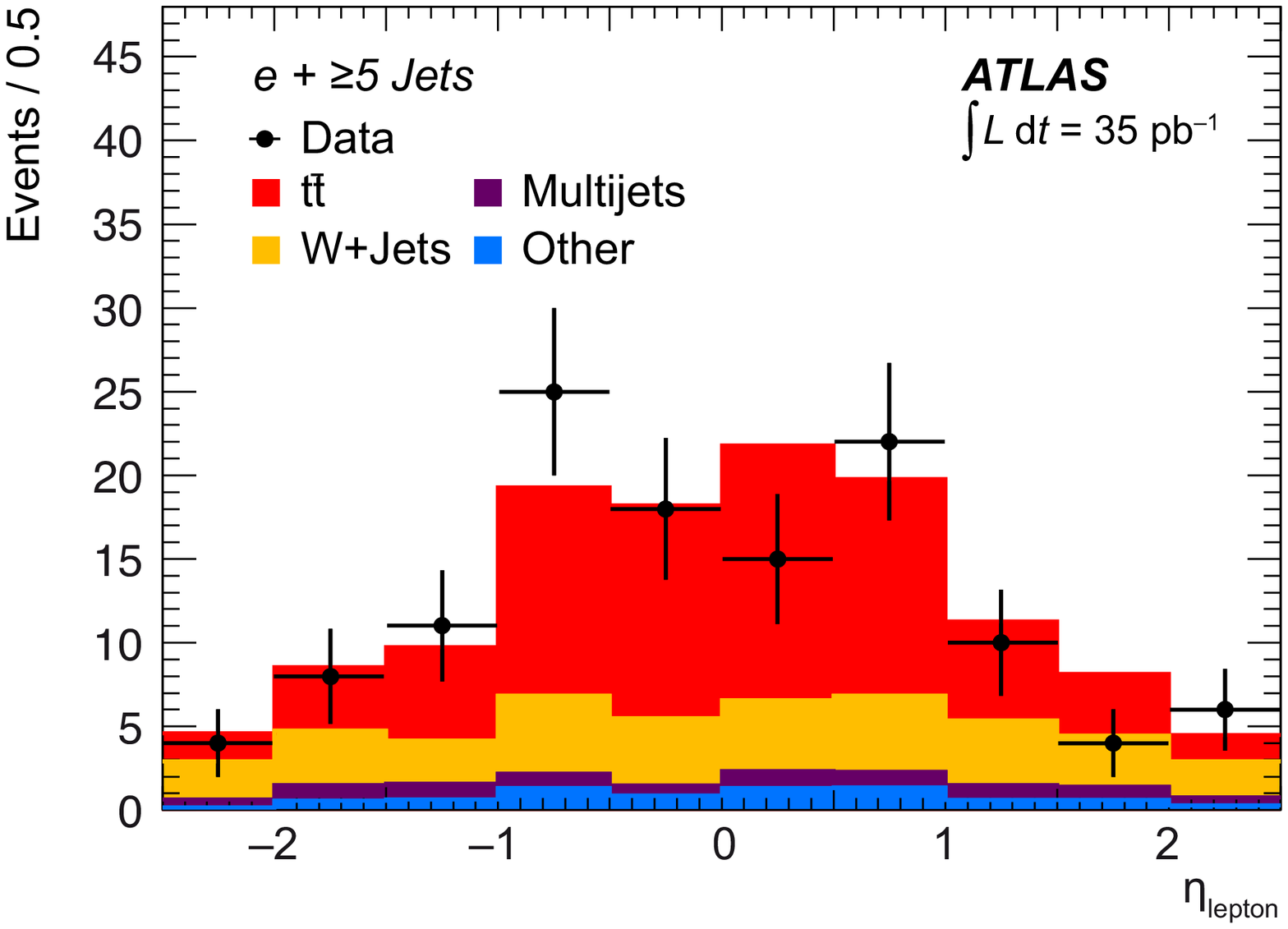} \\
\includegraphics[width=0.85\linewidth]{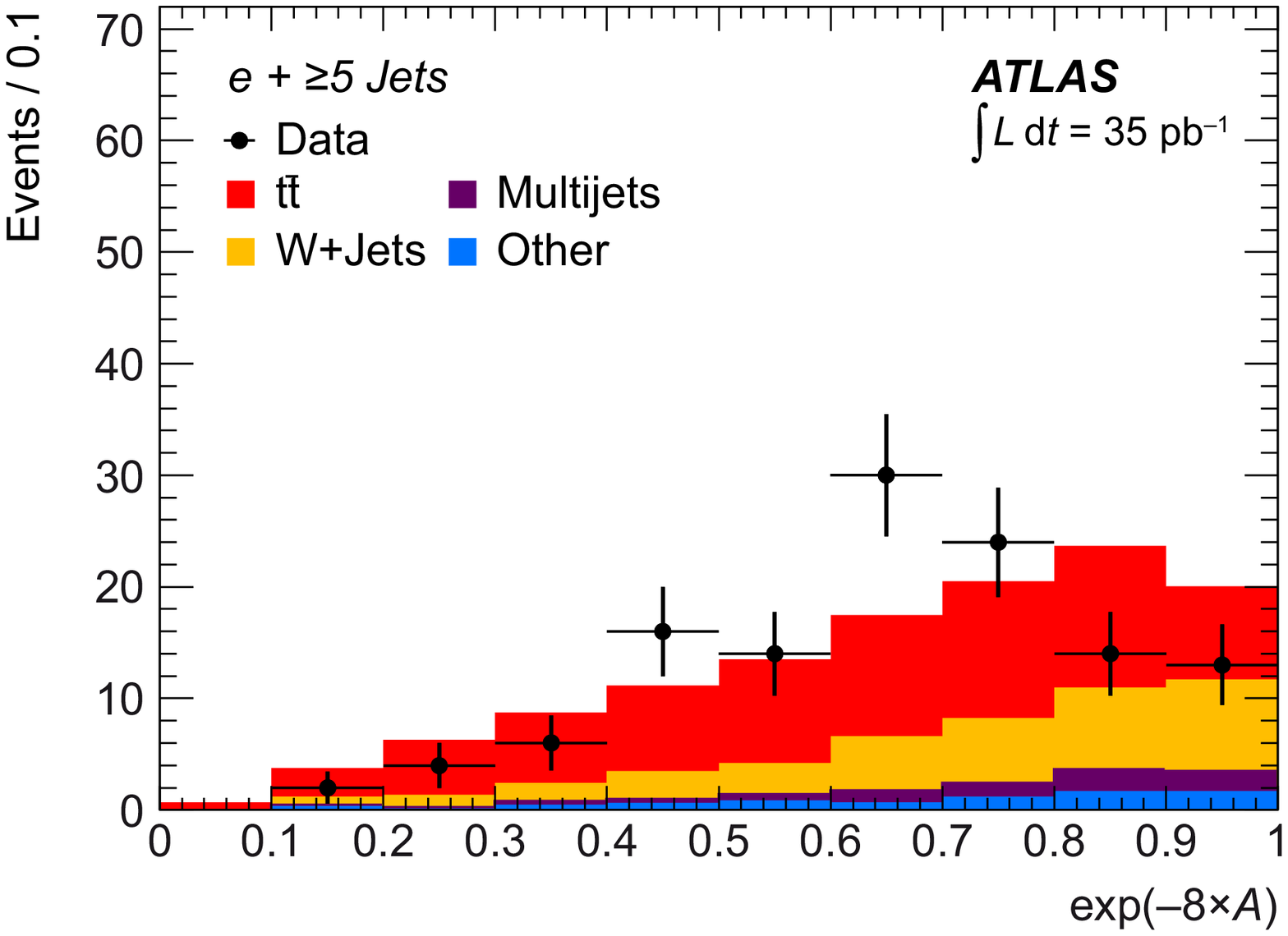} \\
\includegraphics[width=0.85\linewidth]{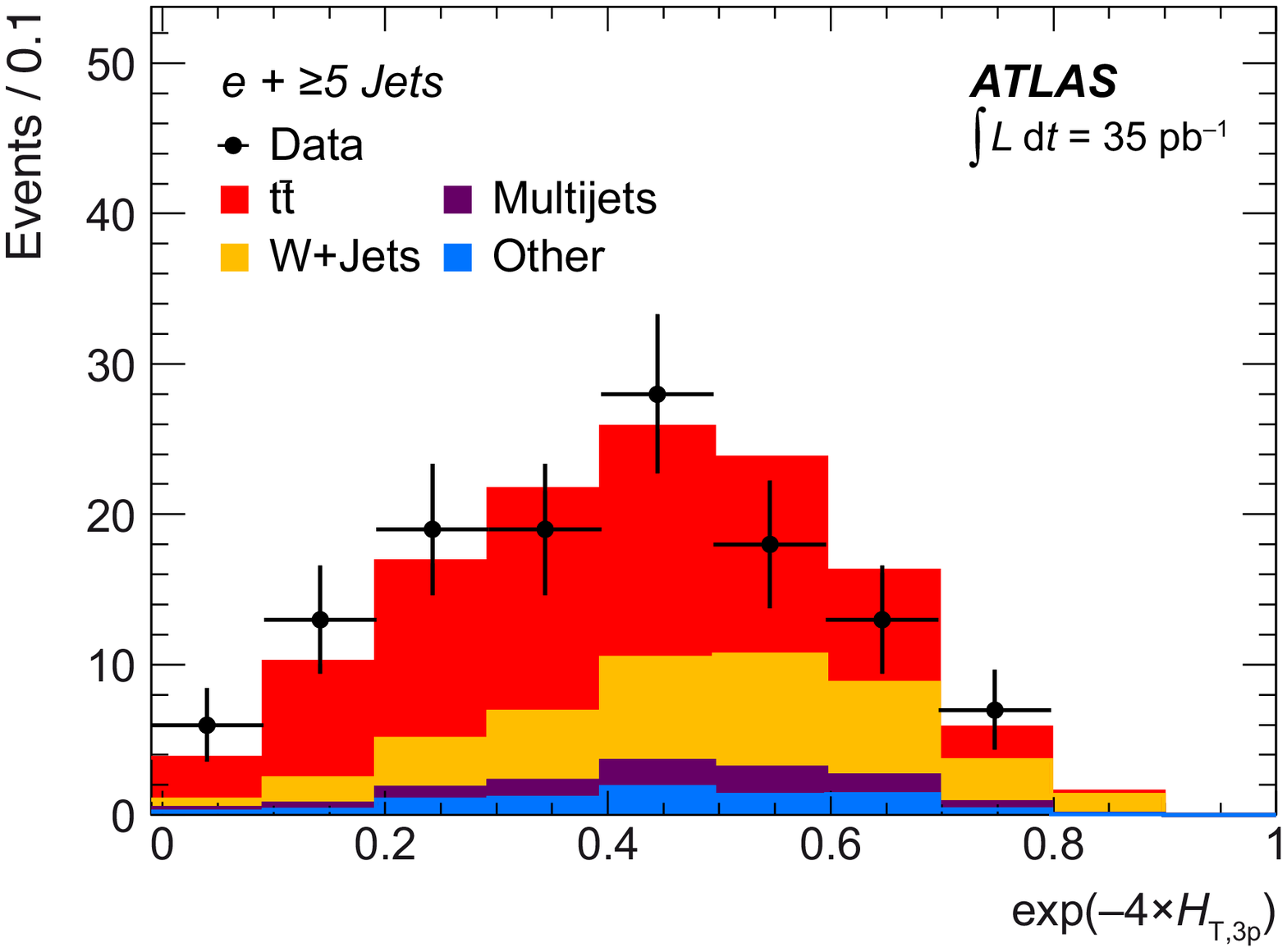} \\
\includegraphics[width=0.85\linewidth]{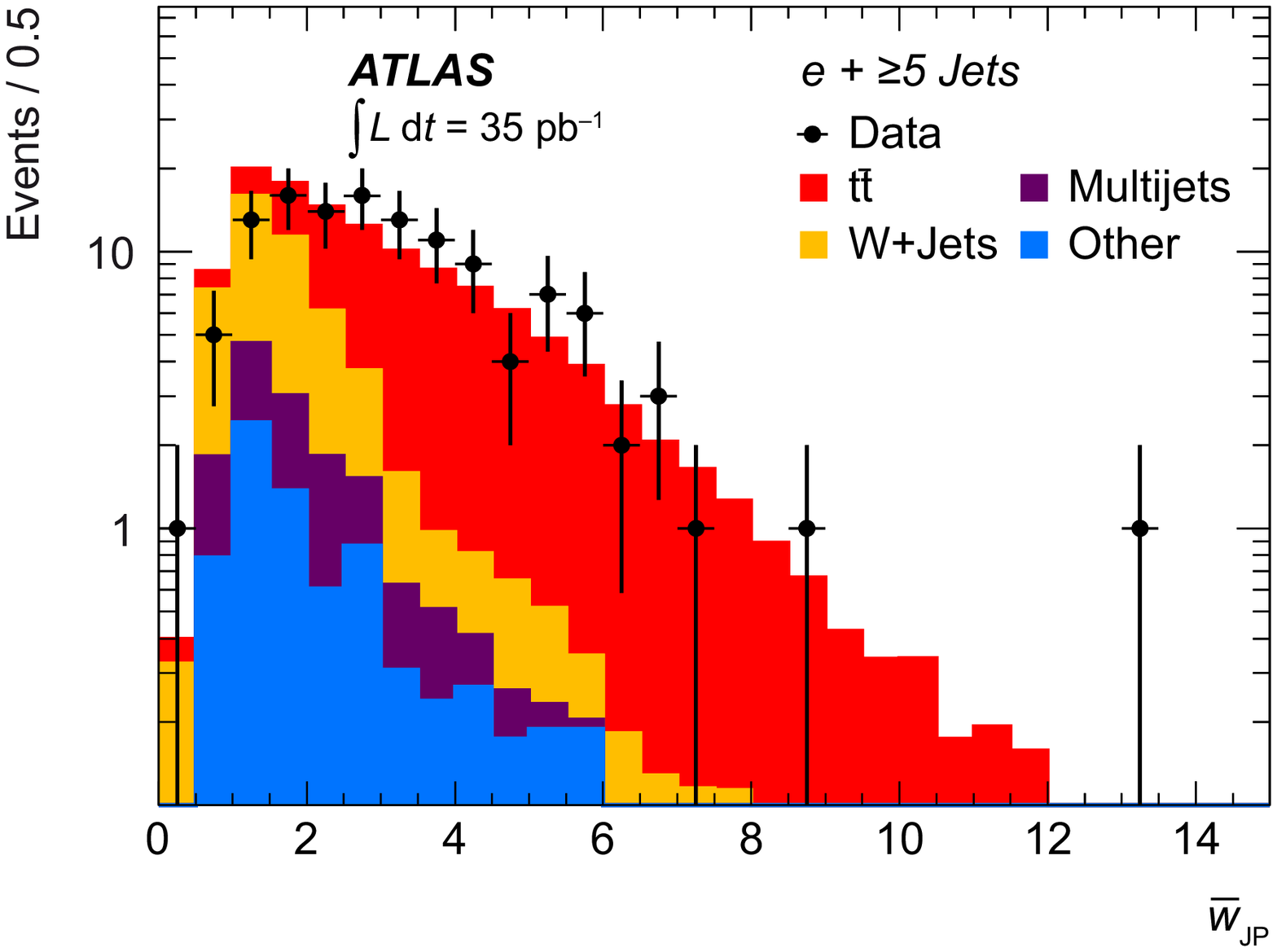} \\
\end{tabular}
\end{center}
\caption{\label{fig:5jetInclusive} 
Input variables to the likelihood discriminants in the inclusive five-jet bin for the electron channel:
lepton $\eta$ (top), 
$\exp(-8 \times \mathcal{A})$ (second from top),
$\exp(-4\times H_{\mathrm{T},3p})$ (third from top) and
${\overline w}_{\mathrm{JP}}$ (bottom).
All simulated processes are normalized to theoretical SM predictions, 
except the multijet background which uses the normalisation
presented in Sec.~\ref{s:back}.
These distributions are used in the tagged analysis.
}
\end{figure}

}
\normalsize

The $\ttbar$ cross-section was extracted by means of a likelihood fit
of the signal and background discriminant distributions to those of the data.
The fit yields the fractions of $\ttbar$ signal and backgrounds in the data sample.
The fit was performed simultaneously to
four samples (three-jet exclusive and four-jet inclusive, electron and muon) in the untagged analysis 
and six samples  (three-jet exclusive, four-jet exclusive and five-jet inclusive, 
electron and muon) in the tagged analysis,
as these were the combinations that provided maximum sensitivity. 
The discriminants were built separately for each jet multiplicity and lepton flavour subsample,
and the different channels were combined in the likelihood fit by multiplying the individual
likelihood functions.

The normalisation of the $\ttbar$ signal templates is the parameter of interest 
in the fit and was allowed to vary freely in both analyses. 
The $\ttbar$ cross-section was assumed to be common to all channels
and the number of $\ttbar$ events in each subsample returned by the fit 
was related to the $\ttbar$ cross-section by the expression
$\sigmattbar = N_{\mathrm sig} / \left ( \int \mathcal L dt \times \epsilon_{\mathrm sig} \right )$, 
where $N_{\mathrm sig}$ is the number of $\ttbar$ events,
$\int \mathcal{L} dt$ is the integrated luminosity and
$\epsilon_{\mathrm sig}$ is the product of the signal acceptance, selection efficiency
and branching ratio, obtained from $\ttbar$ simulation.
The normalisation of the backgrounds was treated differently in the two analyses.
In the untagged analysis the multijet and small backgrounds 
(single-top, diboson and $Z$+jets production) 
were fixed in the fit to their expected contributions, whereas the $W$+jets background was allowed 
to vary freely in each channel.
In the tagged analysis all backgrounds were allowed to vary within the uncertainties 
of their assumed cross-sections, described in Sections~\ref{mc.section}~and~\ref{s:back}.
These uncertainties were used as Gaussian constraints on the cross-section normalisation. 
The robustness of this fitting approach was checked with ensemble tests. 
The central value and uncertainties returned by the fit were shown to be unbiased
for a wide range of input cross-sections.

\section{Systematic uncertainties}
\label{s:systematics}

The evaluation of the systematic uncertainties was performed differently in the two analyses.
The untagged analysis performed pseudo-experiments (PEs) with simulated samples which included 
the various sources of uncertainty. 
For example, for the JES uncertainty, PEs were performed with jet energies scaled up and down 
according to their uncertainties and the impact on the cross-section was evaluated.
The tagged analysis, on the other hand, accounted for most of the changes in the
normalisation and shape of the templates due to systematic uncertainties 
by adding `nuisance' terms to the fit~\cite{profile}.
Templates of the samples with one standard deviation 'up' and 'down' 
variations of the systematic uncertainty source under study were generated 
in addition to the nominal templates. 
The fit interpolated between these templates with a continuous parameter 
by means of a Gaussian constraint.
%Each nuisance parameter is assumed to be Gaussian distributed 
%with mean value zero and unity r.m.s. width. 
Before the fit, the constraint was such that the mean value was zero and the width was one; 
a fitted width less than one means that the data were able to constrain that particular 
source of uncertainty.
%Thus the fitted values of the nuisance parameters correspond to the amount of systematic 
%uncertainty that best fits the data. 
The effects due to 
the modelling of the $W$+jets and multijet background shapes, 
initial and final state radiation,
parton density function of the $\ttbar$ signal,
NLO generator, hadronisation and template statistics
cannot be fully described by a simple linear parameter
controlling the template interpolation.
As a consequence, they were not treated as nuisance terms but
obtained by performing PEs with modified simulated samples,
as was done in the untagged analysis.

The nuisance parameters of the systematic uncertainties were all fitted together
taking into account the correlations among them in the minimisation process. 
As a consequence, the uncertainties on the fitted quantities obtained from the
fit include both the statistical and the total systematic components.
Therefore, to obtain an estimation of the individual contributions to the total uncertainty
in the tagged analysis, each individual systematic uncertainty was obtained 
as the difference in quadrature between the total uncertainty and the uncertainty 
obtained after having fixed the corresponding nuisance parameter to its fitted value.
The central values of the nuisance parameters after the fit agreed with their input values.
The fit was cross-checked using PEs where the starting value of the nuisance parameters
was different than the nominal value. The result was found to be unbiased. 
In addition, large variations of the kinematic dependence of the nuisance
parameters ({\it e.g.} the JES as a function of the jet $\pT$) were considered and
resulted in a negligible impact on the result of the fit.

The systematic uncertainties on the cross-section for both methods are summarised in Table~\ref{tab:sys}. 
The dominant effects in the untagged analysis were JES, multijet and $W$+jets
backgrounds shape and ISR/FSR. The latter was also important in the tagged analysis,
together with the uncertainty related to the signal MC generator.
In addition, this analysis was sensitive to effects related to $b$-tagging,
specifically the determination of the heavy-flavour content of the $W$+jets background 
and the calibration of the $b$-tagging algorithm itself.
The luminosity uncertainty was 3.4\%~\cite{lumiPAPER,lumiCONF}.

\begin{table}[htb]
{\small
\caption{\small
{Statistical and systematic uncertainties on the measured $\ttbar$ 
cross-section in the untagged and tagged analyses. 
Multijet and small backgrounds normalisation uncertainties are already included in the 
statistical uncertainty ({\it a/i}) in the tagged analysis.
$W$+jets heavy-flavour content and $b$-tagging calibration do not apply ({\it n/a}) to the untagged analysis.
The luminosity uncertainty is not included in the table.}
\label{tab:sys}}
%\vspace*{-0.2cm}
\begin{center}
\begin{tabular}{ l  r r  r r}
\hline
\hline
Method                 & \multicolumn{2}{c}{Untagged} & \multicolumn{2}{c}{Tagged}  \\
\hline 
\hline
Statistical Error (\%)                                  & {\bf $+10.1$} & {\bf $-10.1$} & {\bf $+5.8$} & {\bf $-5.7$} \\
\hline
\hline
\multicolumn{3}{l}{Object selection (\%)}  \\
\hline
\hspace*{0.2cm}JES and jet energy resolution            &   $+4.1$   &   $-5.4$   &   $+3.9$   &   $-2.9$       \\
\hspace*{0.2cm}Lepton reconstruction, \\
\hspace*{0.4cm}identification and trigger               &   $+1.7$   &   $-1.6$   &   $+2.1$   &   $-1.8$       \\
\hline
\multicolumn{3}{l}{Background modelling (\%)}\\
\hline
\hspace*{0.2cm}Multijet shape                           &   $+3.5$   &   $-3.5$   &   $+0.8$   &   $-0.8$       \\
\hspace*{0.2cm}Multijet normalisation                   &   $+1.1$   &   $-1.2$   &\multicolumn{2}{c}{\it a/i} \\
%\hspace*{0.2cm}Small backgrounds \\
\hspace*{0.2cm}Small backgrounds norm.                  &   $+0.6$   &   $-0.6$   &\multicolumn{2}{c}{\it a/i} \\
\hspace*{0.2cm}$W$+jets shape                           &   $+3.9$   &   $-3.9$   &   $+1.0$   &   $-1.0$       \\
\hspace*{0.2cm}$W$+jets heavy-flavour content        &\multicolumn{2}{c}{\it n/a} &   $+2.7$   &   $-2.4$       \\
\hspace*{0.2cm}$b$-tagging calibration               &\multicolumn{2}{c}{\it n/a} &   $+4.1$   &   $-3.8$       \\
\hline
\multicolumn{3}{l}{$t \bar t$ signal modelling (\%)}  \\
\hline
\hspace*{0.2cm}ISR/FSR                                  &   $+6.3$   &   $-2.1$   &   $+5.2$   &   $-5.2$       \\
\hspace*{0.2cm}NLO generator                            &   $+3.3$   &   $-3.3$   &   $+4.2$   &   $-4.2$       \\
\hspace*{0.2cm}Hadronisation                            &   $+2.1$   &   $-2.1$   &   $+0.4$   &   $-0.4$       \\
\hspace*{0.2cm}PDF                                      &   $+1.8$   &   $-1.8$   &   $+1.5$   &   $-1.5$       \\
\hline
\multicolumn{3}{l}{Others (\%)}  \\
\hline
\hspace*{0.2cm}Simulation of pile-up                    &   $+1.2$   &   $-1.2$   &\multicolumn{2}{c}{$<0.1$}      \\
\hspace*{0.2cm}Template statistics                      &   $+1.3$   &   $-1.3$   &   $+1.1$   &   $-1.1$       \\
\hline
\hline
Systematic Error (\%)                                   & {\bf $+10.5$} & {\bf $-9.4$} & {\bf $+9.7$} & {\bf $-9.0$}  \\
\hline
\hline
\end{tabular}
\end{center}
}
\end{table}

Several cross-checks of the cross-section measurements were performed.
These included the results of the likelihoods applied to individual lepton channels and
$\ttbar$ cross-section measurements done with simpler and complementary approaches,
including cut-and-count methods and fits to kinematic variables such as the reconstructed
top mass. %\cite{ATLAS_Moriond_1,ATLAS_Moriond_2}.
These cross-checks gave consistent results within the uncertainties.

\section{Results and conclusions}
\label{s:results_conclusions}

The results of the likelihood fits applied to the data are shown in 
Figs.~\ref{fig:finalfit_untagged}~and~\ref{fig:finalfit_tagged},
where the distributions of the discriminants in data are overlaid 
on the fitted discriminant distributions of the signal and backgrounds.
The final measured cross-section results are:
$\sigmattbar = 173 \pm 17 \mathrm{(stat.)} \,^{+18}_{-16} \mathrm{(syst.)} \pm 6 \mathrm{(lumi.) \; pb} =
173 ^{+25}_{-24} \; \mathrm{pb}$ 
in the untagged analysis and
$\sigmattbar = 187 \pm 11 \mathrm{(stat.)} \,^{+18}_{-17} \mathrm{(syst.)} \pm 6 \mathrm{(lumi.) \; pb} =
187 ^{+22} _{-21} \; \mathrm{pb}$
in the tagged analysis. 
The two measurements are in agreement with each other.
The latter has a better a priori sensitivity and thus constitutes the main result of this Letter.
It is the most precise $\ttbar$ cross-section measurement at the LHC published to date and is in
good agreement with the SM prediction calculated at NLO plus next-to-leading-log order
165~$\pmasym{11}{16}$~pb~\cite{topxs1,topxs2,topxs3}.

\scriptsize{
\begin{figure}[htb]
\begin{center}
  \includegraphics[width=0.45\textwidth]{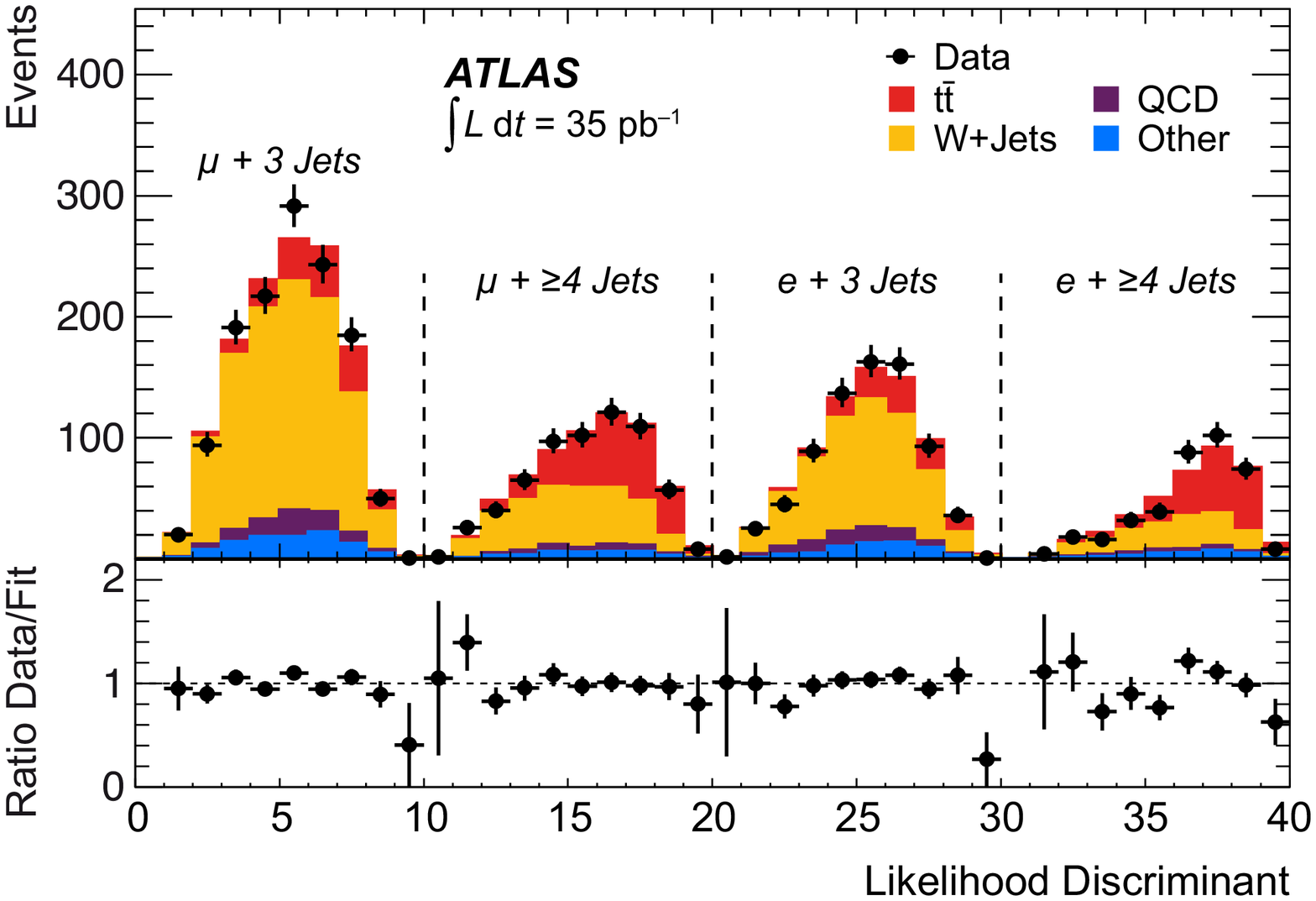} 
\caption{\small {
Untagged analysis:
(Top) The distribution of the likelihood discriminant for data superimposed 
on expectations for signal and backgrounds, scaled to the results of the fit. 
The left bins correspond to the muon channel and the right bins to the electron channel.
(Bottom) The ratio of data to fit result.
}}
\label{fig:finalfit_untagged}
\end{center}
\end{figure}
}

\normalsize

\scriptsize{
\begin{figure}[htb]
\begin{center}
  \includegraphics[width=0.45\textwidth]{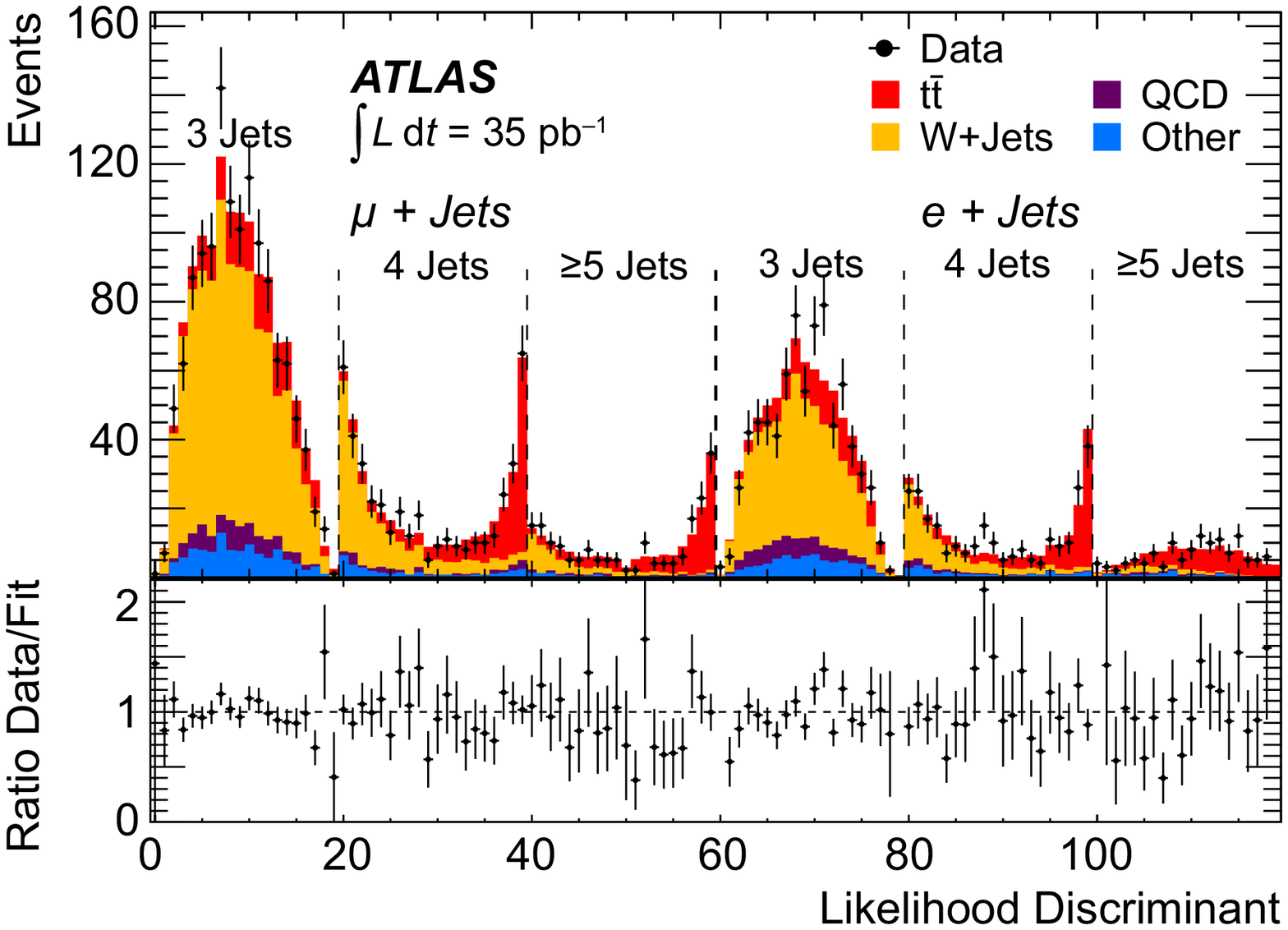} 
\caption{\small {
Tagged analysis:
(Top) The distribution of the likelihood discriminant for data superimposed 
on expectations for signal and backgrounds, scaled to the results of the fit. The left bins correspond to
the muon channel and the right bins to the electron channel.
(Bottom) The ratio of data to fit result.
}}
\label{fig:finalfit_tagged}
\end{center}
\end{figure}
}

\normalsize

\section{Acknowledgements}
\label{s:acknow}

We thank CERN for the very successful operation of the LHC, as well as the
support staff from our institutions without whom ATLAS could not be
operated efficiently.

We acknowledge the support of ANPCyT, Argentina; YerPhI, Armenia; ARC,
Australia; BMWF, Austria; ANAS, Azerbaijan; SSTC, Belarus; CNPq and FAPESP,
Brazil; NSERC, NRC and CFI, Canada; CERN; CONICYT, Chile; CAS, MOST and
NSFC, China; COLCIENCIAS, Colombia; MSMT CR, MPO CR and VSC CR, Czech
Republic; DNRF, DNSRC and Lundbeck Foundation, Denmark; ARTEMIS, European
Union; IN2P3-CNRS, CEA-DSM/IRFU, France; GNAS, Georgia; BMBF, DFG, HGF, MPG
and AvH Foundation, Germany; GSRT, Greece; ISF, MINERVA, GIF, DIP and
Benoziyo Center, Israel; INFN, Italy; MEXT and JSPS, Japan; CNRST, Morocco;
FOM and NWO, Netherlands; RCN, Norway; MNiSW, Poland; GRICES and FCT,
Portugal; MERYS (MECTS), Romania; MES of Russia and ROSATOM, Russian
Federation; JINR; MSTD, Serbia; MSSR, Slovakia; ARRS and MVZT, Slovenia;
DST/NRF, South Africa; MICINN, Spain; SRC and Wallenberg Foundation,
Sweden; SER, SNSF and Cantons of Bern and Geneva, Switzerland; NSC, Taiwan;
TAEK, Turkey; STFC, the Royal Society and Leverhulme Trust, United Kingdom;
DOE and NSF, United States of America.

The crucial computing support from all WLCG partners is acknowledged
gratefully, in particular from CERN and the ATLAS Tier-1 facilities at
TRIUMF (Canada), NDGF (Denmark, Norway, Sweden), CC-IN2P3 (France),
KIT/GridKA (Germany), INFN-CNAF (Italy), NL-T1 (Netherlands), PIC (Spain),
ASGC (Taiwan), RAL (UK) and BNL (USA) and in the Tier-2 facilities
worldwide.

%%%%%%%%%%%%%%%%%%%%%%%%%%%%%%%%%%%%%%%%%%%%%%%%%%%%%%%%%%%%%%%%%%%%%%%%%%%%%%%
% Bibliography
%%%%%%%%%%%%%%%%%%%%%%%%%%%%%%%%%%%%%%%%%%%%%%%%%%%%%%%%%%%%%%%%%%%%%%%%%%%%%%%

\clearpage
\onecolumn
% ATLAS Collaboration author list for 30-SEP-2011
% Data extracted on 16-Mar-2012 for paperid 142
%\documentclass[11pt]{article}
%\usepackage{a4wide}\begin{document}
\begin{flushleft}
{\Large The ATLAS Collaboration}

\bigskip

G.~Aad$^{\rm 48}$,
B.~Abbott$^{\rm 110}$,
J.~Abdallah$^{\rm 11}$,
A.A.~Abdelalim$^{\rm 49}$,
A.~Abdesselam$^{\rm 117}$,
O.~Abdinov$^{\rm 10}$,
B.~Abi$^{\rm 111}$,
M.~Abolins$^{\rm 87}$,
H.~Abramowicz$^{\rm 152}$,
H.~Abreu$^{\rm 114}$
% ...
\bigskip

% ...
$^{10}$ Institute of Physics, Azerbaijan Academy of Sciences, Baku, Azerbaijan\\
$^{11}$ Institut de F\'isica d'Altes Energies and Departament de F\'isica de la Universitat Aut\`onoma  de Barcelona and ICREA, Barcelona, Spain\\
% ...
$^{48}$ Fakult\"{a}t f\"{u}r Mathematik und Physik, Albert-Ludwigs-Universit\"{a}t, Freiburg i.Br., Germany\\
$^{49}$ Section de Physique, Universit\'e de Gen\`eve, Geneva, Switzerland\\
% ...
$^{87}$ Department of Physics and Astronomy, Michigan State University, East Lansing MI, United States of America\\
% ...
$^{110}$ Homer L. Dodge Department of Physics and Astronomy, University of Oklahoma, Norman OK, United States of America\\
$^{111}$ Department of Physics, Oklahoma State University, Stillwater OK, United States of America\\
% ...
$^{114}$ LAL, Univ. Paris-Sud and CNRS/IN2P3, Orsay, France\\
% ...
$^{117}$ Department of Physics, Oxford University, Oxford, United Kingdom\\
% ...
$^{152}$ Raymond and Beverly Sackler School of Physics and Astronomy, Tel Aviv University, Tel Aviv, Israel\\
% ...
\end{flushleft}

%\end{document} 

\end{document}